%% file: summative-eval-paper.tex
\documentclass[manuscript, screen]{acmart}

\input{macros.tex}

\def\plaintitle{\rTisane: Externalizing conceptual models for data analysis increases engagement with domain knowledge and improves statistical model quality}

\def\plainauthor{Eunice Jun, Edward Misback, Jeffrey Heer, Ren{\'e} Just}

\def\plainkeywords{statistical analysis; linear modeling; end-user programming; end-user elicitation; domain-specific language}

\makeatletter
\def\url@leostyle{%
  \@ifundefined{selectfont}{
    \def\UrlFont{\sf}
  }{
    \def\UrlFont{\small\bf\ttfamily}
  }}
\makeatother
\def\pprw{8.5in}
\def\pprh{11in}

\definecolor{linkColor}{RGB}{6,125,233}
\hypersetup{%
  pdftitle={\plaintitle},
 pdfauthor={\plainauthor},
  pdfkeywords={\plainkeywords},
  pdfdisplaydoctitle=true, 
  bookmarksnumbered,
  pdfstartview={FitH},
  colorlinks,
  citecolor=black,
  filecolor=black,
  linkcolor=black,
  urlcolor=linkColor,
  breaklinks=true,
  hypertexnames=false
}

\author{Eunice Jun}
\email{emjun@cs.ucla.edu}
\orcid{0000-0002-4050-4284}
\affiliation{%
  \institution{University of California, Los Angeles and University of Washington}
  \country{USA}
}

\author{Edward Misback}
\email{misback@cs.washington.edu}
\orcid{0009-0003-9474-0826}
\affiliation{%
  \institution{University of Washington}
  \country{USA}
}

\author{Jeffrey Heer}
\email{jheer@cs.washington.edu}
\orcid{0000-0002-6175-1655}
\affiliation{%
  \institution{University of Washington}
  \country{USA}
}

\author{Ren{\'e} Just}
\email{rjust@cs.washington.edu}
\orcid{}
\affiliation{%
 \institution{University of Washington}
 \country{USA}
 }

\renewcommand{\shortauthors}{Jun et al.}

\hypersetup{draft}
\begin{document}

\title[Externalizing conceptual models increases engagement with domain knowledge, improves statistical model quality]{\plaintitle}
\input{abstract.tex}

\maketitle





\input{figures_tables.tex}

\input{intro.tex}
\input{related_work.tex}

\input{exploratory.tex}
\input{rTisane.tex}
\input{lab-study.tex}

\input{discussion.tex}
\input{limitations.tex}

\input{conclusion.tex}


\typeout{}
\bibliographystyle{SIGCHI-Reference-Format}
\bibliography{references}

\end{document}

%% file: macros.tex
\usepackage{tabularx} 
\usepackage{booktabs} 
\usepackage[ruled]{algorithm2e} 
\usepackage[colorinlistoftodos]{todonotes}
\usepackage{verbatim}
\usepackage{graphicx}
\usepackage{xspace}

\usepackage{balance}       
\usepackage[T1]{fontenc}   
\usepackage{color}
\usepackage{booktabs}

\usepackage{microtype}        
\usepackage{ccicons}          

\usepackage{cuted}
\usepackage{capt-of}

\usepackage{upquote}
\usepackage{listings}
\usepackage{xcolor}
\usepackage{float}

\definecolor{codegreen}{rgb}{0,0.6,0}
\definecolor{codegray}{rgb}{0.5,0.5,0.5}
\definecolor{codepurple}{rgb}{0.58,0,0.82}

\lstdefinestyle{mystyle}{
    backgroundcolor=\color{white},
    commentstyle=\color{codegreen},
    keywordstyle=\color{magenta},
    numberstyle=\tiny\color{codegray},
    stringstyle=\color{codepurple},
    basicstyle=\ttfamily\footnotesize,
    breakatwhitespace=false,
    breaklines=true,
    captionpos=b,
    keepspaces=true,
    numbers=left,
    numbersep=5pt,
    showspaces=false,
    showstringspaces=false,
    showtabs=false,
    tabsize=2
}

\definecolor{light-gray}{gray}{0.95}

\definecolor{lightgray}{rgb}{0.83, 0.83, 0.83}

\usepackage{multirow}
\usepackage[normalem]{ulem}

\usepackage{blindtext}
\usepackage{subcaption}
\usepackage{caption}

\usepackage{enumitem}
\usepackage{graphicx}
\usepackage{multicol}

\usepackage{appendix}

\usepackage{tikz}
\usetikzlibrary{positioning,arrows.meta,graphs,backgrounds,fit,calc,quotes,shapes.multipart}
\usepackage{pgfplots}
\usepackage{pgfplotstable}

\pgfplotsset{width=3.5cm,
    tick label style={font=\footnotesize}
}

\definecolor{tisanecodetop}{RGB}{74,63,106}
\definecolor{closecolor}{RGB}{225,93,87}
\definecolor{minimizecolor}{RGB}{245,215,90}
\definecolor{maximizecolor}{RGB}{111,208,74}

\SetAlFnt{\small}
\SetAlCapFnt{\small}
\SetAlCapNameFnt{\small}
\SetAlCapHSkip{0pt}
\IncMargin{-\parindent}

\newcommand{\shortquote}[1]{``\emph{#1}''}
\newcommand{\longquote}[1]{\vspace{-1pt}\begin{quote}``\emph{#1}''\end{quote}}
\newcommand{\longquoteEmphAdded}[1]{\vspace{-1pt}\begin{quote}``\emph{#1}'' (emphasis added).\end{quote}}
\newcommand{\theme}[1]{\vspace{-1pt}\subsubsection{#1} }

\def\lme{\texttt{lme4}\xspace}

\def\dataSet{dataset\xspace}
\def\tisane{Tisane\xspace}
\def\SDSLlong{DSL\xspace}
\def\SDSL{DSL\xspace}
\def\rTisane{rTisane\xspace}
\def\rTisanes{rTisane's\xspace}

\def\evalConceptualModels{\textbf{RQ1 - Conceptual model specification}\xspace}
\def\evalStatisticalModels{\textbf{RQ2 - Statistical model quality}\xspace}

\usepackage{setspace}

\newcommand\colH[1]{\multicolumn{1}{c}{\textbf{#1}}}

\newcolumntype{H}{>{\setbox0=\hbox\bgroup}c<{\egroup}@{}}

\newcommand{\assume}[1]{\tikz[baseline]\node[circle,inner xsep=2pt,inner ysep=0pt,
        draw=black,fill=white,scale=0.8,anchor=base] {#1};}

\def\<#1>{\codeid{#1}}
\newcommand{\codeid}[1]{\ifmmode{\mbox{\small\ttfamily{#1}}}\else{\small\ttfamily #1}\fi}
\newcommand{\codeidsmall}[1]{\ifmmode{\mbox{\smaller\ttfamily{#1}}}\else{\smaller\ttfamily #1}\fi}

%% file: abstract.tex
\begin{abstract}

    Statistical models should accurately reflect analysts' domain knowledge
    about variables and their relationships. While recent tools let analysts
    express these assumptions and use them to produce a resulting statistical
    model, it remains unclear what analysts want to express and how
    externalization impacts statistical model quality. This paper addresses
    these gaps. We first conduct an exploratory study of analysts using a
    domain-specific language (DSL) to express \textit{conceptual models}. We observe a
    preference for detailing \textit{how} variables relate and a desire to allow, and
    then later resolve, ambiguity in their conceptual models. We leverage these
    findings to develop \rTisane, a DSL for expressing conceptual models
    augmented with an interactive disambiguation process. In a controlled
    evaluation, we find that \rTisane's DSL helps analysts engage more deeply
    with and accurately externalize their assumptions. \rTisane also leads to
    statistical models that match analysts' assumptions, maintain analysis
    intent, and better fit the data.

\end{abstract}

%% file: figures_tables.tex
\lstdefinestyle{rtisanestyle}{
    backgroundcolor=\color{white},
    commentstyle=\color{codegreen},
    keywordstyle=\color{magenta}, 
    numberstyle=\tiny\color{codegray},
    stringstyle=\color{codepurple},
    basicstyle=\ttfamily\footnotesize,
    breakatwhitespace=false,
    breaklines=true,
    captionpos=b,
    keepspaces=true,
    numbers=left,
    numbersep=5pt,
    showspaces=false,
    showstringspaces=false,
    showtabs=false,
    tabsize=2,
    alsoletter={/, <-},
    deletekeywords={library, cm, order, list, /, <-},
    morekeywords={Unit, continuous, categories, ConceptualModel, assume, hypothesize, causes, relates, interacts, query},
    literate={/}{{/}}1 {<-}{{<-}}2
}

\newcommand{\rTisaneProgram}{
    \begin{figure} [tb]
    \lstinputlisting[
        language={R}, 
        style=rtisanestyle,
        caption={\textbf{Sample \rTisane program adapted from P8 in the evaluation study.} When declaring variables (lines 3-18), specifying cardinality is optional with data.
        Executing this program opens up the conceptual model disambiguation interface in \autoref{fig:figureConceptualModelsDisambiguation}.},
        label={lst:rTisaneProgram}
    ]{figures/rTisane-example.R}
    \end{figure}
}

\newcommand{\conceptualModelsScaffold}{
    \begin{figure}[htbp]
        \centering
        \includegraphics[width=.75\linewidth]{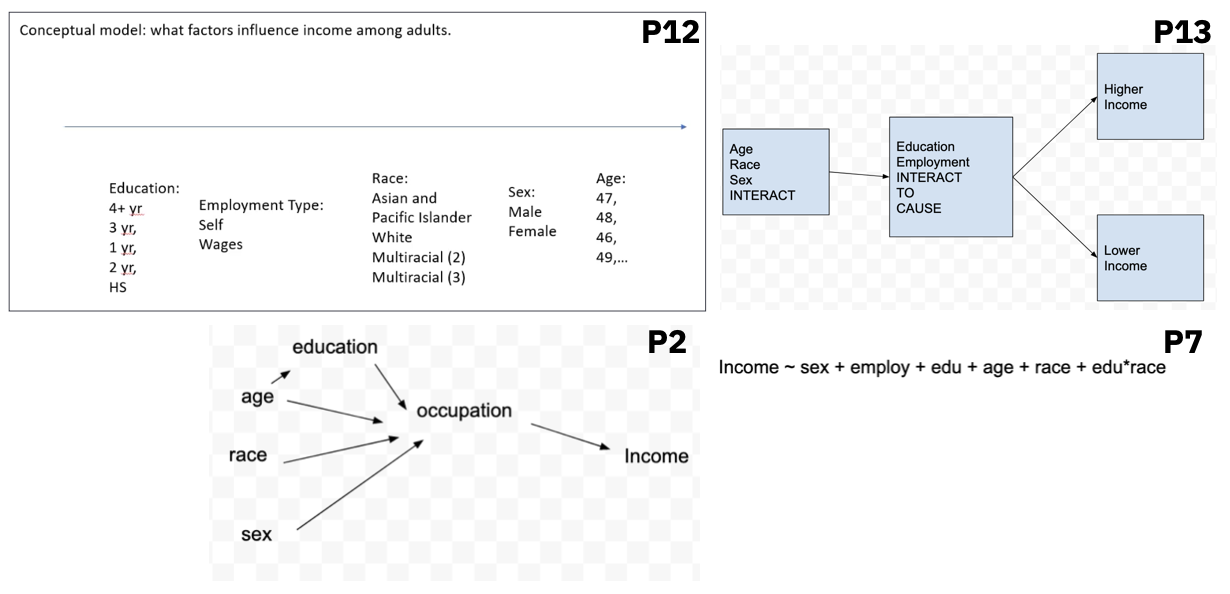}
        \caption{\textbf{Evaluation: Example conceptual models without \rTisane.}}
            \begin{small}
            \begin{minipage}{\linewidth}
                Participants expressed conceptual
                models without \rTisane in a plurality of formats, including in
                natural language, a timeline [P12], graphs [P2, P13], and
                directly as a statistical model [P7]. 
                Using \rTisane, participants were able to express their conceptual models in a more structured way, which promoted deeper reflection on assumptions and consideration of additional relationships. 
            \end{minipage}
            \end{small}
        \label{fig:figureConceptualModelsScaffold}
    \end{figure}
}

\newcommand{\conceptualModelDisambiguation}{
    \begin{figure}[h]
        \centering
        \includegraphics[width=.95\linewidth]{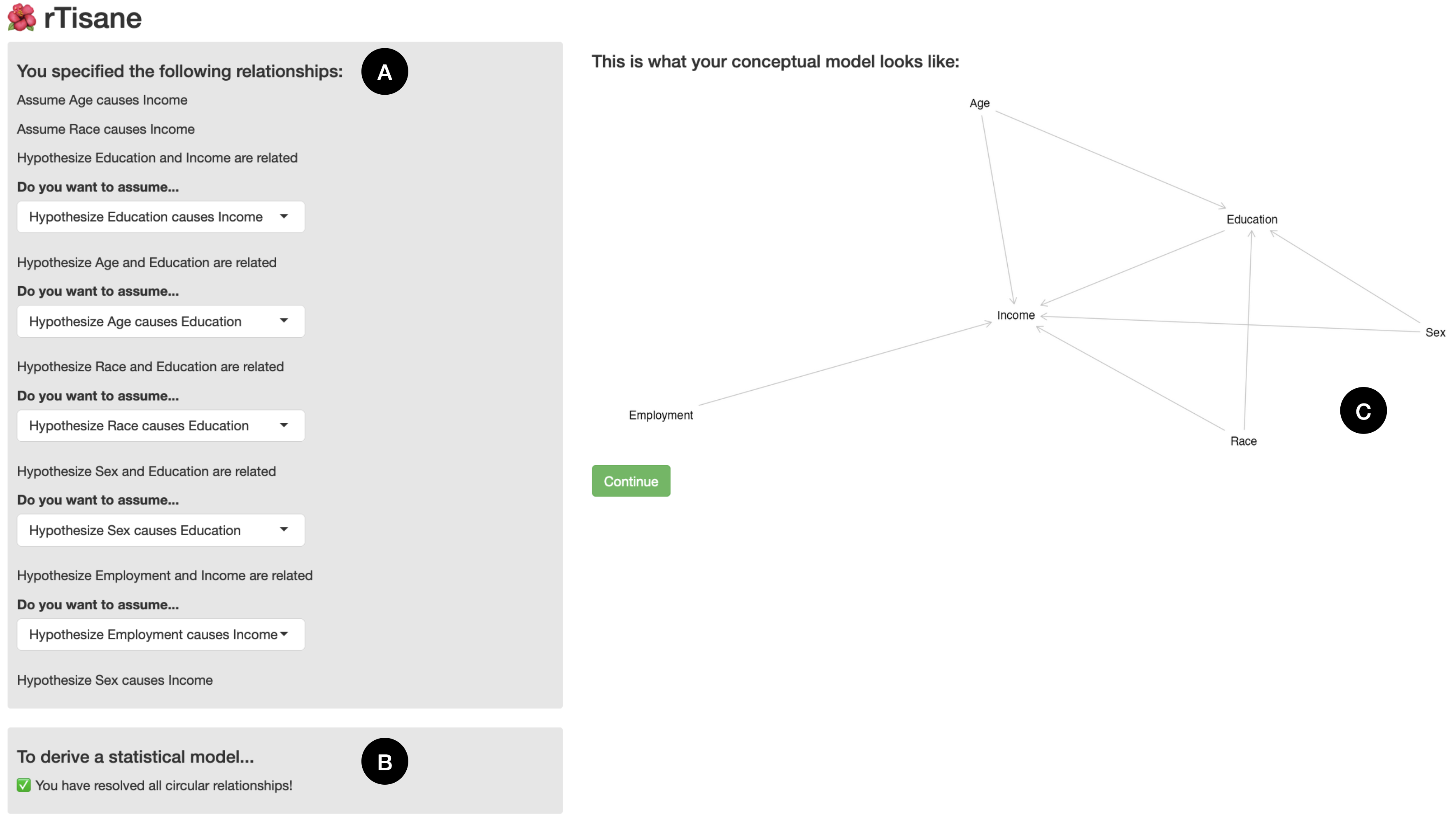}
        \caption{\textbf{\rTisanes conceptual model disambiguation interface.}}
            \begin{small}
            \begin{minipage}{\linewidth}
                Upon executing the example program in \autoref{lst:rTisaneProgram}, analysts see the above interface. 
                To answer the query and derive a statistical model from a conceptual model, \rTisane has analysts clarify and confirm their conceptual model. 
                (A) The side panel shows options for resolving ambiguities in the conceptual model due to \relates relationships (lines 24-28 in \autoref{lst:rTisaneProgram}).
                (B) \rTisane checks and follows up with questions about breaking any cycles that hinder statistical model derivation. 
                (C) The interface visualizes the underlying graph, updating as analysts resolve ambiguities and break cycles.
                Upon hitting the continue button, analysts see the statistical model disambiguation interface in \autoref{fig:figureStatisticalModelsDisambiguation}.
            \end{minipage}
            \end{small}
        \label{fig:figureConceptualModelsDisambiguation}
    \end{figure}
}

\newcommand{\statisticalModelDisambiguation}{
    \begin{figure}[b!]
        \centering
        \includegraphics[width=.95\linewidth]{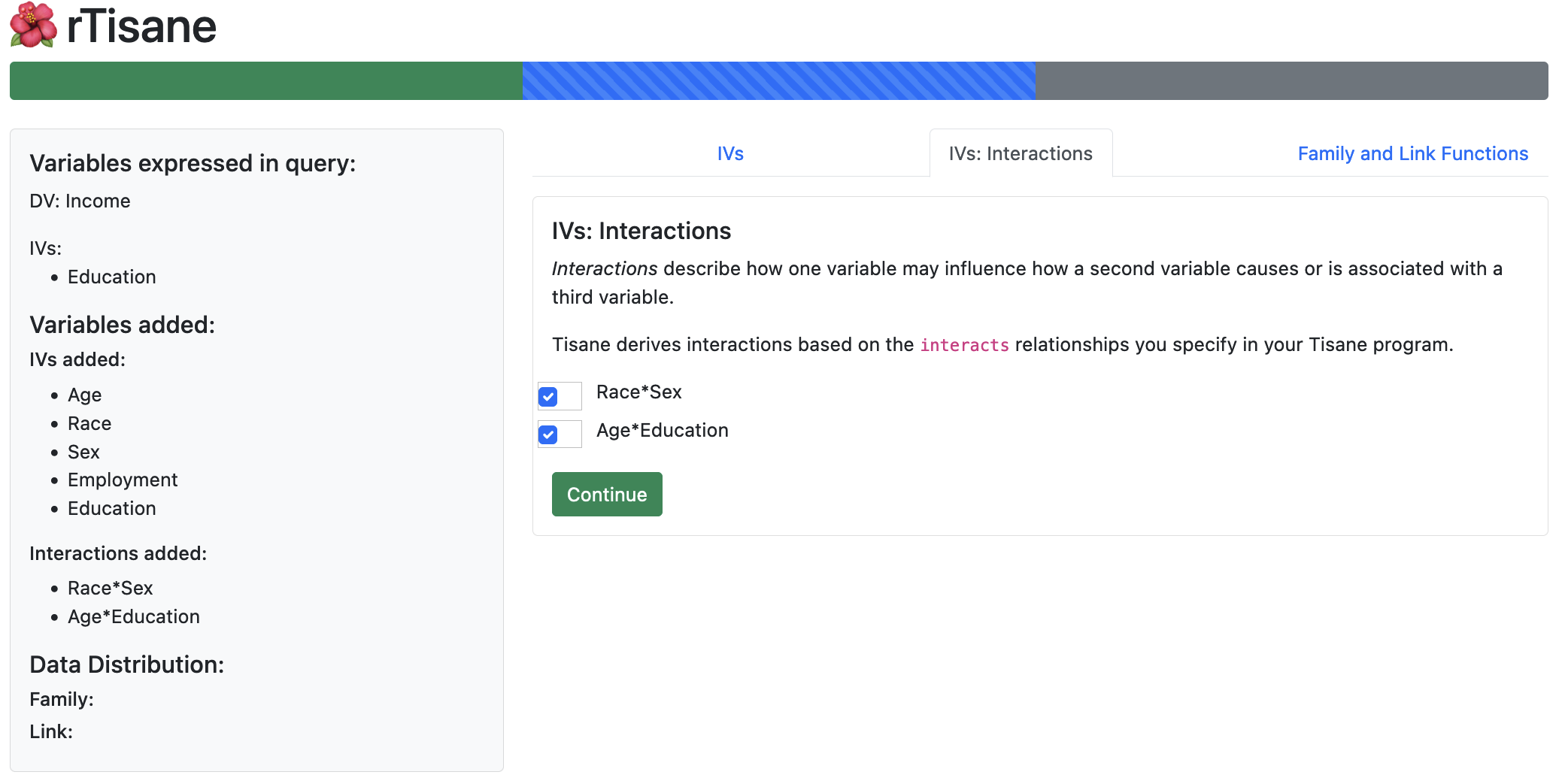}
        \caption{\textbf{\rTisanes statistical model disambiguation interface.}}
            \begin{small}
            \begin{minipage}{\linewidth}
                \rTisane shows an interface explaining automatic statistical modeling decisions. 
                \rTisane also asks analysts questions to narrow the space of possible
                statistical models to a final one. 
                Statistical model
                disambiguation occurs after conceptual
                model disambiguation (\autoref{fig:figureConceptualModelsDisambiguation}).
            \end{minipage}
            \end{small}
        \label{fig:figureStatisticalModelsDisambiguation}
    \end{figure}
}

\newcommand{\mysmaller}{\fontsize{9pt}{9pt}\selectfont}
\newcommand{\smalltt}[1]{{\mysmaller \texttt{#1}}}

\newcommand{\tableSummativeEvalParticipants}{
  \begin{table}[h]
        \caption{\textbf{Evaluation participants.}}
        \begin{small}
        \begin{minipage}{\linewidth}
            Participants came from a diversity of fields and job roles. All self-reported having familiarity with generalized linear models, experience programming in R, and significant data analysis experience. \\\vspace{-5pt}
        \end{minipage}
        \end{small}
        \centering
        \begin{tabular}{l|l|l}
            \toprule
            \colH{ID} & \colH{Field} & \colH{Role} \\ 
            \midrule
            P1 & Statistics & Data Scientist \\ 
            P2 & Mechanical Engineering & Graduate Student \\ 
            P3 & Data Science & Research Assistant \\ 
            P4 & Political Science & Data Science Educator \\ 
            P5 & Data Science & Professor \\ 
            P6 & Biology & Visiting Scientist \\ 
            P7 & Psychology & Quantitative User Researcher \\ 
            P8 & Bioinformatics  & Researcher  \\ 
            P9 & Data Analytics & Senior Operations Data Analyst \\ 
            P10 & Automotive Engineering & PhD Student \\ 
            P11 & Data Analysis & Research Analyst \\ 
            P12 & Data Analytics & Data Engineer \\ 
            P13 & Public Health & Data Scientist \\ 
        \end{tabular}
    \label{tab:summativeEvaluationParticipants}
  \end{table}
}

\newcommand{\tableModelScores}{
    \renewcommand{\arraystretch}{0.3}
    \begin{table}
        \caption{\textbf{Evaluation: Comparing statistical models authored with and without \rTisane.}}
        \begin{footnotesize}
        \begin{minipage}{\linewidth}
            Using \rTisane, analysts authored statistical models that fit the data just as well or better than without \rTisane. 
            For each participant, the better AIC and BIC scores are in bold. AIC and BIC measure how well a statistical model fits data, with lower scores indicating better fit.
            P7, P9, P11, and P13 authored statistical models with \rTisane first, as indicated by \textsuperscript{1}. 
            P2's statistical model without tool support fits the data better in part because he prioritized data fit at the expense of maintaining analysis intent and fidelity to his conceptual model (see \autoref{themeChangeAnalysisIntent} for more details).
            For P7, P8, and P13, there are no bold scores because the statistical models with and without \rTisane are identical. 
            We did not observe a difference in statistical model quality depending on tool support order, except in the case of P11. 
            When asked to author a statistical model without \rTisane, P11 took the output model from \rTisane, deemed poor model fit based on the AIC score, log transformed \texttt{Income}, and then fit the revised model as their own. 
            To perform the log transform, P11 dropped observations where \texttt{Income$=$0}, explaining the marked difference in AIC/BIC scores between tool support conditions, as indicated by \textsuperscript{a}. 
        \end{minipage}
        \end{footnotesize}
        \centering
        \mysmaller
        \begin{tabularx}{\linewidth}{p{.025\linewidth} p{.055\linewidth} >{\raggedright}p{0.53\linewidth} r r r}
            \toprule
            \colH{ID} & \colH{Tool} & \colH{Statistical model} & \colH{df} & \colH{AIC} & \colH{BIC} \\
            \midrule \\ 
            P2                       
            & None       
            & \smalltt{lm(data\$Income $\sim$ data\$Employment + data\$Age + data\$Race + data\$Education + data\$Sex + data\$Age*data\$Employment + data\$Race*data\$Employment + data\$Education*data\$Employment + data\$Sex*data\$Employment)}
            & 37 & \textbf{60,327,741} 
            & \textbf{60,328,211} \\
            & & & & \\ 
            & \rTisane   
            & \smalltt{glm(formula=Income $\sim$ Employment, family=gaussian(link='identity'), data=data)}
            & 4 & 60,781,341           
            & 60,781,392 \\
            \midrule \\ 
            P4                       & None       & \smalltt{lm(Income $\sim$ Age + Education + Employment + Race + Sex, data=data)}                                                                                                                                                  & 15 & 60,358,715           & 60,358,906          \\
            & & & & \\ 
                                     & \rTisane   & \smalltt{glm(formula=Income $\sim$ Education + Age + Education*Sex + Employment + Race + Sex, family=gaussian(link='identity'), data=data)}                                                                                                 & 19 & \textbf{60,332,919} & \textbf{60,333,161 } \\
            \midrule \\ 
            P7\textsuperscript{1}                       & None       & \smalltt{lm(formula = Income $\sim$ Age + Race + Education + Employment + Sex, data = data)}                                                                                                                                      & 15 & 60,358,715          & 60,358,906           \\
            & & & & \\ 
                                    & \rTisane   & \smalltt{glm(formula=Income $\sim$ Sex + Age + Employment + Race + Education, family=gaussian(link='identity'), data=data)}                                                                                                               & 15 & 60,358,715           & 60,358,906          \\
            \midrule \\ 
            P8                       & None       & \smalltt{lm(Income $\sim$ Sex*Race + Employment + Education + Race*Sex + Age, data = data)}                                                                                                                                       & 20 & 60,354,038          & 60,354,292          \\
            & & & & \\ 
            & \rTisane   & \smalltt{glm(formula=Income $\sim$ Age + Race*Sex + Employment + Age*Education, family=gaussian(link='identity'), data=data)}                                                                                                           & 24 & \textbf{60,351,454} & \textbf{60,351,759} \\
            \midrule \\ 
            P9\textsuperscript{1}                       & None       & \smalltt{smf.OLS("Income $\sim$ Age + C(Race) + C(Education) + C(Employment) + C(Sex)", data=df)}                                                                                                                                 & 15 & 60,358,715          & 60,358,906           \\
                                     & \rTisane   & \smalltt{glm(formula=Income $\sim$ Employment + Race + Sex + Education + Age, family=gaussian(link='identity'), data=data)}                                                                                                               & 15 & 60,358,715          & 60,358,906         \\
            \midrule \\ 
            P10                      & None       & \smalltt{sm.OLS.from\_formula("Income $\sim$ Age", data=df)}                                                                                                                                                                     & 3 & 60,876,872          & 60,876,910          \\
            & & & & \\ 
            & \rTisane   & \smalltt{glm(formula=Income $\sim$ Employment + Sex + Education + Age + Sex*Education, family=gaussian(link='identity'), data=data)}                                                                                                      & 14 & \textbf{60,339,137} & \textbf{60,339,315} \\
            \midrule \\ 
            P11\textsuperscript{1}                      & None       & \smalltt{glm(log\_income $\sim$ Employment + Race + Age + Education + Sex, family = "gaussian", data=data)}                                                                                                                       & 15 & \textbf{11,741,899\textsuperscript{a}}        & \textbf{11,742,089\textsuperscript{a}}          \\
            & & & & \\ 
            & \rTisane   & \smalltt{glm(formula=Income $\sim$ Employment + Race + Sex + Education + Age, family=gaussian(link='identity'), data=data)}                                                                                                               & 15 & 60,358,715          & 60,358,906          \\
            \midrule \\ 
            P13\textsuperscript{1}                      & None       & \smalltt{glm(Income $\sim$ Age*factor(Race)*factor(Sex) + factor(Education)*factor(Employment), family="gaussian", data)}                                                                                                           & 39 & 60,331,749          & 60,332,244          \\
            & & & & \\ 
            & \rTisane   & \smalltt{glm(formula=Income $\sim$ Employment + Age*Race*Sex + Education*Employment + Education, family=gaussian(link='identity'), data=data)}                                                                                          & 39 & 60,331,749          & 60,332,244         
        \end{tabularx}
        \label{tab:statisticalModelComparison}
    \end{table}
}

\newcommand{\exampleConceptualModel}{
    \begin{figure}[htbp]
        \centering
        \includegraphics[width=.5\linewidth]{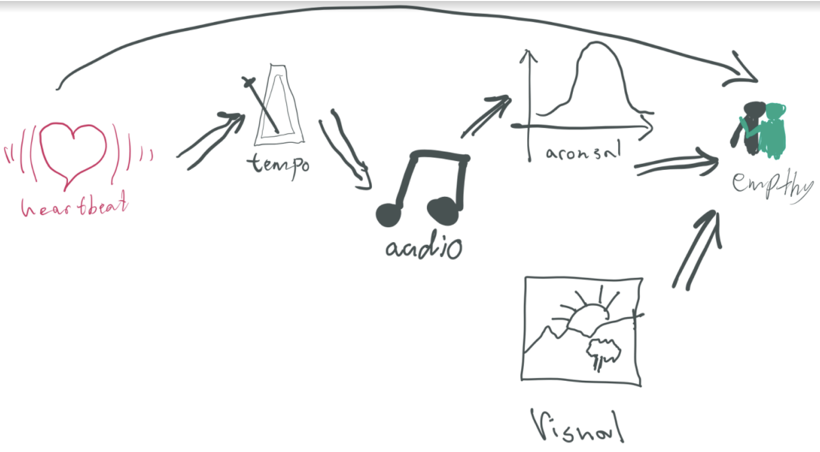}
        \caption{\textbf{Example conceptual model from exploratory study.}}
            \begin{small}
            \begin{minipage}{\linewidth}
                In the exploratory study, participants expressed their conceptual models on their own before coming to the lab. 
                In the lab, they used \tisane, an open-source DSL, to specify their conceptual models. 
                We observed how analysts wanted to express their assumptions, including what challenges they faced, as described in~\autoref{sec:exploratoryStudy}. 
                Study findings informed our design goals for \rTisane (\autoref{sec:rtisane_design_implications}).
            \end{minipage}
            \end{small}
        \label{fig:figureExampleConceptualModel}
    \end{figure}
}

\newcommand{\paperOverview}{
    \begin{figure}[t!]
        \centering
        \includegraphics[width=1.\linewidth]{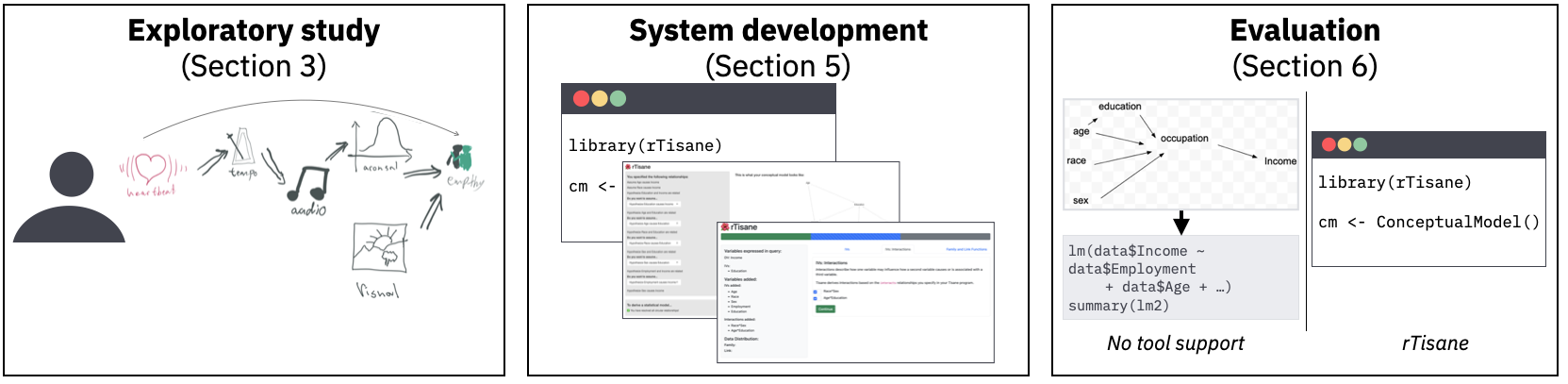}
        \caption{\textbf{Visual overview of paper.}}
            \begin{small}
            \begin{minipage}{\linewidth}
                Through an exploratory study, we investigate how to better support statistical non-experts in specifying their conceptual models (\autoref{sec:exploratoryStudy}). 
                Based on findings, we develop \rTisane, a system for specifying and refining conceptual models in order to derive statistical models (\autoref{sec:rTisane}). 
                We compare \rTisane to a scaffolded workflow in a within-subjects controlled lab study (\autoref{sec:summativeEval}). 
                We find that using \rTisane to externalize conceptual models deepens consideration of implicit assumptions and supports accurate explication of conceptual models.
                We also find that \rTisane enables analysts to author statistical models that represent their assumptions, maintain their analysis intent, and fit the data better. 
            \end{minipage}
            \end{small}
        \label{fig:figurePaperOverview}
    \end{figure}
}

\newcommand{\rTisaneOverview}{
    \begin{figure}[t!]
        \centering
        \includegraphics[width=1.\linewidth]{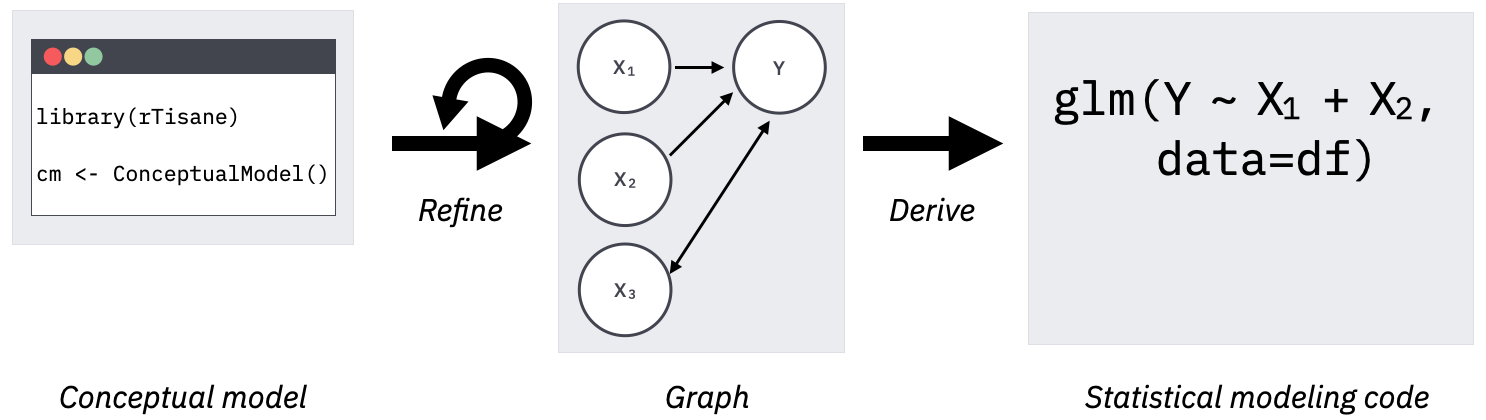}
        \caption{\textbf{Overview of \rTisane.}}
            \begin{small}
            \begin{minipage}{\linewidth}
                \rTisane provides a DSL for specifying conceptual models (left box).
                Analysts verify and refine their conceptual models as the first step of a two-phase interactive disambiguation process (left arrow, see \autoref{fig:figureConceptualModelsDisambiguation}). 
                Interactive refinement updates the internal graph representation (middle box).
                \rTisane traverses this graph to formulate possible statistical models (right arrow, see \autoref{fig:figureStatisticalModelsDisambiguation}).
                Analysts learn about \rTisanes modeling decisions and can change them prior to getting a statistical modeling script as output (right box).
            \end{minipage}
            \end{small}
        \label{fig:rTisaneOverview}
    \end{figure}
}

\newcommand{\evaluationFlow}{
    \begin{figure}[t!]
        \centering
        \includegraphics[width=.85\linewidth]{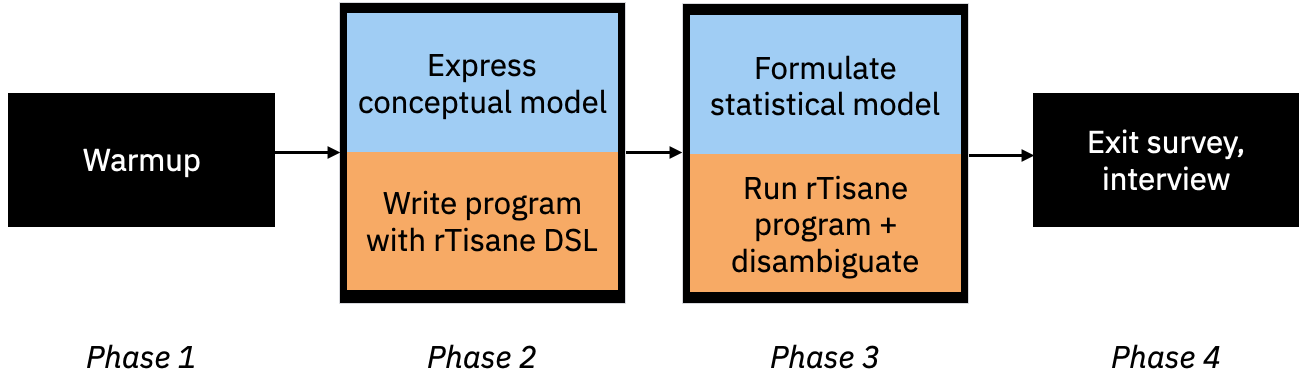}
        \caption{\textbf{Evaluation phases and conditions.}}
            \begin{small}
            \begin{minipage}{\linewidth}
                We conducted a within-subjects controlled lab study where we compared \rTisane to a scaffolded workflow without tool support (2 conditions). 
                All participants completed four phases: warm-up, conceptual model specification, statistical model formulation, and exit survey with interview.
                For Phases 2 and 3, participants either completed the task (i) without tool support (blue) then with \rTisane (orange) or (ii) with \rTisane (orange) then without tool support (blue). 
                Each participant saw the same condition order in Phases 2 and 3.
            \end{minipage}
            \end{small}
        \label{fig:evaluationFlow}
    \end{figure}
}

%% file: intro.tex
\section{Introduction}

In order to answer research questions and test hypotheses,
analysts must translate their research questions and hypotheses into statistical
models. To do so accurately, analysts need to reflect on their implicit
understanding of the domain and consider how to represent this conceptual
knowledge in a statistical model. For example, consider a health policy
researcher interested in accurately estimating the influence of insurance coverage on
health outcomes. To formulate a statistical model, they consider prior work on
how insurance coverage, race, education, and health outcomes relate to each
other and other constructs. Then, they go to formulate a statistical model
including or excluding covariates to account for confounding in these
relationships~\cite{cinelli2020controls}. 
A researcher who skips this process may overlook relevant conceptual relationships or implicit assumptions, resulting in statistical
models (and conclusions) that are faulty or meaningless as answers to their motivating research question.

Key to this explanatory modeling process is analysts' domain knowledge, captured in
\textit{process models}~\cite{mcelreath2020statistical} or \textit{conceptual
models}~\cite{jun2022hypoForm}. Conceptual models include variables and their relationships that are important
to a domain. \autoref{fig:figurePaperOverview} shows an example conceptual model from our exploratory study (\autoref{sec:exploratoryStudy}). A number of software tools exist for building conceptual models. For example,
\tisane~\cite{jun2022tisane}, an open-source library for authoring generalized
linear models with or without mixed effects, enables analysts to explicate their
conceptual models and derives valid statistical models from them. \tisane has
helped HCI researchers catch and fix analysis bugs prior to
publication~\cite{baughan2022dissociation}. Other tools such as
Dagitty~\cite{textor2011dagitty} and DoWhy~\cite{sharma2020dowhy} also support
analysts in externalizing conceptual models as causal graphs
to reason through statistical modeling choices. These software tools 
support (i) conceptual model specification and (ii) statistical model
formulation based on expressed conceptual models. 

\paperOverview

To benefit from these tools, analysts must be able to accurately externalize their
implicit conceptual models (goal (i)). This goal presents two usability challenges. First,
tools should make it easy for analysts to express their conceptual models.
At the very least, tools should not hinder specification. 
Second, analysts need guidance on which implicit assumptions are important to
externalize. Addressing both challenges is particularly important for making
these analysis tools usable for domain experts who have statistical experience
but limited expertise (i.e., many researchers). 

After analysts externalize conceptual models, tools must formulate statistical models (goal (ii)) in order to obtain high-quality statistical inferences.
To ensure quality, there are two challenges to statistical model formulation: 
fidelity of the statistical model to the conceptual model
and good statistical model fit. 
These criteria provide checks on one another. 
For instance, for any data set, an overfit statistical model can be found that satisfies 
the model fit criterion as well as possible without accurately representing the 
analyst’s implicit conceptual model. As another
example, a statistical model representing an unreasonable conceptual model may not fit real-world data well. 
We prioritize correspondence of conceptual models to statistical models and 
then, given this correspondence, maximize model fit.


This paper investigates how to support both accurate conceptual model
specification and quality statistical model formulation. We focus on the design
and implementation of a domain-specific language (DSL) for expressing conceptual
models and using conceptual models to author statistical models. We focus on DSL design since
end-users and graphical systems alike can benefit from DSLs. Our
users are statistical non-experts who have domain expertise, experience with
generalized linear modeling, and experience programming in R, but are not
statistical experts. We refer to these end-users as \textit{statistical
non-experts}. 

We start with an exploratory study to identify challenges statistical
non-experts face when expressing their conceptual models. We find that analysts
want to specify \textit{how} variables relate causally (e.g., ``more heartbeat
alignment leads to more empathy'') instead of stating \textit{that} one causes
another (e.g., ``heartbeat alignment causes empathy''). Analysts also want to
express ambiguity in their conceptual models, and, if necessary to derive
statistical models, clarify any ambiguity in an interactive refinement step.
Based on these findings, we develop \rTisane, a system for externalizing
conceptual models to author generalized linear models (GLMs). \rTisane consists of (i)
a DSL for expressing conceptual models and (ii) a two-phase interactive
disambiguation process for refining conceptual models and then deriving statistical
models. \rTisane leverages an informative graphical user interface (GUI) for disambiguation.
The result of this entire process is a script for fitting a
statistical model that is guaranteed to reflect the expressed-then-refined
conceptual model. To assess the impact of \rTisane on conceptual model
specification and statistical model formulation, we compare \rTisane to a
scaffolded workflow without tool support in a within-subjects lab study. We find that \rTisanes DSL
makes it easy for analysts to specify conceptual models and guides them to think
more critically about their implicit assumptions. Furthermore, \rTisane helps
analysts focus on their analysis intents, and analysts are not surprised by
\rTisanes output statistical models. Using \rTisane, analysts are able to
author statistical models that fit the data just as well, if not better, than
statistical models they author without tool support. \autoref{fig:figurePaperOverview} visually shows the three parts of this paper.


In summary, we contribute
\begin{itemize}
    \item A study identifying how statistical non-experts want and are capable
    of expressing their implicit domain assumptions,
    \item The open-source \rTisane system, which provides new language constructs for
    expressing conceptual models and a two-phase interactive disambiguation process for
    resolving ambiguity in conceptual models and deriving statistical models, and 
    \item Evidence from a controlled lab study about how tool
    support for externalizing conceptual models to author statistical models leads to thorough conceptual model specification and higher quality statistical models.
\end{itemize}

%% file: related_work.tex
\section{Background and Related Work}
We contextualize our work on \rTisane in relation to empirical studies and
theories of data analysis, tools for conceptual modeling, and tools for
authoring statistical analyses. 

\subsection{Empirical studies and theories of data analysis}

Data analysis is an iterative process of data discovery, wrangling, profiling,
modeling, and reporting~\cite{kandel2012enterprise}. Exploratory data analysis
helps analysts refine their data, analysis goals, and
hypotheses~\cite{alspaugh2018futzing,wongsuphasawat2019EDAgoals,battle2019EVA}.
Following exploration, analysts want to probe into relationships between
variables in their data through statistical models. Statistical modeling
involves considering numerous analysis decisions and choosing among a range of
analysis alternatives. Liu, Althoff, and Heer~\cite{liu2019paths} identified
numerous decision points throughout the data lifecycle, which they call
\textit{end-to-end analysis}. They found that analysts often revisit key
decisions during data collection, wrangling, modeling, and evaluation. Liu,
Althoff, and Heer also found that researchers executed and selectively reported
analyses that were already found in prior work and familiar to the research
community. Furthermore, Liu, Boukhelifa, and Eagan~\cite{liu2019understanding}
group analysis alternatives into cognitive (e.g., shifts in conceptual
hypotheses), artifact (e.g., choice in statistical tools), and execution (e.g.,
computational tuning) \textit{levels of abstraction}. \textit{Cognitive}
alternatives involve more conceptual shifts and changes (e.g., mental models,
hypotheses). \textit{Artifact} alternatives pertain to tooling (e.g., which
software is used for analysis?), model (e.g., what is the general mathematical
approach?), and data choices (e.g., which dataset is used?). \textit{Execution}
alternatives are closely related to artifact alternatives but are more
fine-grained programmatic decisions (e.g., hyperparameter tuning).

Jun et al.'s conceptual framework of \textit{hypothesis
formalization}~\cite{jun2021hypothesisFormalization} encompasses all three
levels of abstraction and describes more granularly how these levels cooperate
with one another. Hypothesis formalization is the process by which analysts
translate their research questions and hypotheses into statistical models. To
craft statistical model programs, analysts incorporate and refine their domain
knowledge, study design, statistical modeling choices, and computational
instantiations of statistical models. Central to hypothesis formalization is the
connection between implicit domain assumptions and a statistical model
implementation. Implicit assumptions are encoded in informal \textit{conceptual
models}, or \textit{process models}~\cite{mcelreath2020statistical}. This paper
focuses on how to provide tool support for analysts to externalize, iterate on, and
formalize their implicit conceptual models. The resulting system, \rTisane,
facilitates one pass of hypothesis formalization in a potentially iterative
modeling workflow (e.g., Bayesian Workflow~\cite{gelman2020modelExpansion}).

Furthermore, Grolemund and Wickham argue for statistical data analysis as a
sensemaking activity~\cite{grolemund2014cognitive}. Building upon the importance
of \textit{external representations} in Russell et al.'s theory of
sensemaking~\cite{russell1993cost}, Grolemund and Wickham argue for the
importance of representing and re-representing conceptual knowledge in a
\textit{schema}. Conceptual models are the external representations, or schema,
this paper focuses on. We show how DSL primitives and interactive disambiguation
can support conceptual modeling and how appropriate support ultimately
facilitates sensemaking during and after statistical data
analysis~\cite{grolemund2014cognitive}.

\subsection{Tools for conceptual modeling}
Despite the centrality of conceptual modeling to hypothesis formalization, few
tools to support this step exist. Analysts can use general purpose text editing
applications (e.g., Google Docs, Microsoft Word), whiteboards (e.g., manual or
online), and diagramming software (e.g., Figma, Keynote) to document and share
their implicit conceptual models. While usable, these software tools do not
scaffold the conceptual modeling process so that it can lead to statistical
models. On the other hand, tools such as Dagitty~\cite{textor2011dagitty},
CausalWizard~\cite{causalWizard}, and DoWhy~\cite{sharma2020dowhy} help analysts specify causal diagrams
and calculate causal estimands. Yet, these tools are designed for statistical
experts who are comfortable expressing causal diagrams directly.

In this paper, we ask how we might design for both usability and rigor in
expressing conceptual models. Through an iterative design process with
statistical non-experts, we develop \rTisane with the aim to ease conceptual
modeling and reify the connection between conceptual and statistical models for
both statistical non-experts and experts.  

\subsection{Tools for authoring statistical analyses}
There is a vibrant ecosystem of tool support for statistical analysis. Libraries
in programming languages such as Python, R, and Julia~\cite{bezanson2017julia}
support a wide range of analyses. Tools such as JMP~\cite{jmp}, SAS~\cite{sas}, and
SPSS~\cite{spss} do not require programming and provide graphical user interfaces
for selecting and executing statistical analysis approaches. However, existing
software tools prioritize mathematical expressivity and computational control
over explicit support for translating research questions and hypotheses into
statistical analyses~\cite{jun2022hypoForm}. In fact, none elicit conceptual
models to seed the statistical authoring process. 

Researchers have proposed new DSLs and approaches that use explicit
specifications of implicit conceptual assumptions to derive valid analyses. For
instance, using Tea~\cite{jun2019tea}, analysts express hypotheses and study
designs and rely on the system to automatically infer and execute a set of
valid Null Hypothesis Significance Tests. Furthermore,
Tisane~\cite{jun2022tisane} is a mixed-initiative system for authoring
generalized linear models with or without mixed effects. Tisane provides a study
design specification language for expressing conceptual and data relationships
between variables and derives statistical models based on these. In this work,
we use Tisane's open-source
implementation\footnote{\url{https://github.com/emjun/tisane}} to design a study
investigating challenges analysts face when expressing their implicit domain
assumptions. We use Tisane because its implementation is publicly available, it
is the first system to bridge conceptual and statistical modeling, and our focus
is on how to best support conceptual modeling during analysis.
Furthermore, while case studies of Tisane validated the feasibility and
desirability of using conceptual models to author statistical
models~\cite{jun2022tisane}, the lab study in this paper delves deeper into how
and why using conceptual models to author statistical analyses is beneficial.
Finally, while the new DSL we design and evalute, \rTisane, is scoped to output
only generalized linear models, our findings generalize to primitives in Tisane
and other systems (e.g., DoWhy~\cite{sharma2020dowhy}),
that could result in more complex
statistical models.

%% file: exploratory.tex
\def\unit{\texttt{Unit}\xspace}
\def\measure{\texttt{Measure}\xspace}
\def\setup{\texttt{SetUp}\xspace}

\section{Exploratory lab study} \label{sec:exploratoryStudy}
We aimed to understand the ways in which statistical non-experts want to
articulate their implicit domain knowledge. 
We used an existing open-source library, \tisane~\cite{jun2022tisane}, to probe
into analysts' internal processes\footnote{For the lab study, we re-implemented
\tisane (originally in Python) in R due to R's widespread adoption in data
science and use in the research methods course from which we recruited
participants.}. This approach helped us articulate design goals for developing
\rTisane (\autoref{sec:rtisane_design_implications}). 


\subsection{Method}
We recruited participants through a graduate-level quantitative research methods
course as a convenience sample. This allowed us to control recent exposure to
statistical concepts. Five computer science PhD students volunteered to participate.

The study consisted of two parts: (i) a take-home assignment and (ii) an in-lab
session. The take-home assignment asked participants to read a recently
published CHI paper~\cite{winters2021heartbeat}\footnote{We chose the specific
paper because (i) we believed its topic (i.e., biosignals and empathy) would be
broadly approachable and (ii) the statistical methods the authors used (i.e., generalized linear
models) would be familiar with students enrolled in the research methods
course and aligned with our research goals.} and describe the paper's research questions and hypotheses, the
authors' conceptual models, the study's design, and ways to analyze the data to
answer the research questions. We designed the assignment to ensure that
participants engaged with the paper's key ideas before coming into the lab. The
researcher reviewed each submission to prepare participant-specific questions
for a semi-structured, think-aloud lab session. 


At the start of the lab session, participants reviewed their homework submissions
to remind themselves of the paper. The paper and participants' homework
responses remained available for reference throughout the study. Then,
participants completed three tasks: (i) declaring variables, (ii) specifying
study designs, and (iii) expressing conceptual models. For each task,
participants started with \tisane's language constructs to express their intent
and discussed their confusions, how they understood each presented construct,
and what they wanted to specify but could not (if applicable). The researcher
repeatedly reminded participants that the constructs presented were prototype
possibilities and that expressing their intentions was more important than using
the constructs or getting the syntax correct. Throughout, the researcher paid
particular attention to where \tisane broke down for participants and asked
follow-up questions to probe deeper into why. The researcher considered such
breakdowns as openings into semantic mismatches between the end-user and the
DSL. 

We iteratively coded homework submissions, audio transcripts from the lab
sessions, and participants' artifacts from the lab studies. We also consulted the researcher's detailed
notes from the lab sessions. 


\subsection{Key Observations}
All participants demonstrated a working knowledge of the assigned paper's
motivating research questions, study design, and general study procedure. 
We made the following four key observations about how statistical non-experts want
to express their conceptual models. 
Based on these observations, we derived design goals for \rTisane
(\autoref{sec:rtisane_design_implications}).

\subsubsection{Analysts want to express how variables relate to one another in detail.}
Analysts have an intuitive understanding of causality but bluntly stating that a
variable causes another does not capture the richness or nuance of their
implicit domain knowledge. Additional annotations about how a variable
influences another are necessary.

When defining ``causes,'' P2 described \shortquote{[Causes] is...like when we teach
logic...it's like implication, right?....So I'm saying if we are observing an
emotion and...emotion observed can lead to a change in emotional perspective.}
P0, P1, and P3 contrasted a bidirectional relationship between variables,
encapsulated in the \texttt{associates\_with} construct in \tisane, to
their implicit understanding of ``causes.'' For instance, P1 stated \shortquote{the most
like, utilitarian definition by if A causes B, then by changing A, I can change
B whereas \texttt{associates\_with} means that...if I can turn dial A, B might
not change.} In addition to differentiating between causal and associative
relationships, three participants [P0, P1, P3] provided statements of
\emph{specifically how} a variable influenced another in the conceptual models
submitted as homework. For example, P0 wrote, \shortquote{Hearing a heartbeat that seems
to be aligned with visual cues makes someone feel \emph{more strongly} what
another person is feeling} (emphasis added), specifying a \emph{positive influence} of
``hearing a heartbeat'' on empathy. 



\subsubsection{Analysts find moderation difficult to separate from bivariate relationships.}
Participants consistently found \tisane's \texttt{moderates}
construct difficult to understand [P0, P1, P2, P3]. 
This construct is used to specify when one or more variables affect the strength or direction of the influence an independent variable has on a dependent variable.
Participants expressed
confusion about what moderation implied about the relationship between
two variables. For example, P3 grappled with if \texttt{moderates} was shorthand for
expressing associative relationships between each independent variable and the
dependent variable, how moderation implies causal relationships, and if
statistical and conceptual definitions of moderation differed from each other:
\longquote{[L]et's say there's two independent variables and one dependent variable. And
each of the [independent] variables individually is not correlated with the
outcome. But if you put them together, then the correlation appears....I mean,
it's sort of a philosophical question of whether, like each of the ones
individually causes [the dependent variable] in that case. But thinking from
a...statistical perspective, I think that's a situation where you might be able
to express...language and experience level together cause lines of code but
individually they don't because no individual correlation would appear there.}
Therefore, a clear delineation between bivariate relationships and partial statistical specifications of interaction terms is necessary. 

\subsubsection{Analysts distinguish between known and suspected relationships.} \label{subsubsec:knownVsuspected}
Participants described relationships established in prior work as
``assumptions'' or ``assertions'' to check separately from the key research
questions that tested ``suspected'' relationships. P0 described how \longquoteEmphAdded{maybe we
have to differentiate as to like the \emph{known} [relationships] are kind of
the things you're \emph{assuming} there's relationships between these things
whereas the \emph{suspected}...[are] the things kind of like your research
questions are saying like, `We \emph{think} there's this relationship
but...it's what we're testing for'} Similarly, P4 suggested
that Tisane should warn end-users when assumptions about known relationships are
violated in a given data set: \longquoteEmphAdded{I would also say that it would be very handy to
be able to say, kind of \emph{assert} that language has no effect on the line
of code. And be warned if it's not the case, like if your \emph{assertion} is
not...verified automatically with the DSL, but warned...that while your
\emph{assumption} is not holding there is actually an effect, which could be
very handy on your study} The inability to indicate
relationships that are either known or suspected in \tisane may explain why
analysts repeatedly preferred less technical verbs, such as ``influences'' [P0]
or ``leads to'' [P3]. For instance, P0 explained how she preferred
``influences'' over ``causes'' because \shortquote{I guess it's like \emph{a level of
sureness} in it in which, like, `cause' feels more confident in your answers
than `influences'} (emphasis added). Providing a way to label conceptual
relationships as assumptions or the focus of the present analysis could 
make conceptual modeling more approachable and lead to conceptual models that better capture analysts' implicit assumptions.


\subsubsection{Analysts want to consider alternative conceptual structures.}
Participants grappled with what specific structures in a conceptual model meant.
P1 and P3 described how a bidirectional relationship between two variables was
really due to hidden, confounding variables causing both variables. P3 described
how \shortquote{in the real world...when these bidirectional things happen, it means
there's sort of this middleman complex system. Or some like underlying process
of which [two variables are] both components...} Another participant, P2,
wondered aloud about how even what appears to be a direct relationship, may
actually be a chain of indirect or mediated relationships at a lower
granularity: \shortquote{It's like Google Maps. If you zoom out enough, that arrow becomes
a direct arrow.} These observations suggest that while participants can deeply
reflect on what could be happening between variables conceptually, they need
help exploring and figuring out which of these structures matches their implicit
understanding. In other words, analysts need a way to indicate ambiguity about
relationships they can then later re-consider with tool assistance. 
\section{Design Goals} \label{sec:rtisane_design_implications} 

Based on our lab study observations, we derived four design goals to more
accurately capture analysts' conceptual assumptions: 

\def\optionalSpecificity{\textit{DG1 - Optional specificity}\xspace}
\def\interactionAsPartialSpec{\textit{DG2 - Interactions as partial specifications}\xspace}
\def\assumeHypothesize{\textit{DG3 - Distinction between assumed and hypothesized relationships}\xspace}
\def\considerPossibilities{\textit{DG4 - Consideration of possibilities}\xspace}
\begin{itemize}
    \item \optionalSpecificity: Analysts should be able to provide optional
    details about how variables change in relation to each other (e.g., positive
    or negative changes in values) when describing conceptual relationships.
    \item \interactionAsPartialSpec: Analysts should annotate conceptual models with interaction terms they want to include in an output statistical model. 
    \item \assumeHypothesize: Analysts should be able to distinguish between assumed and hypothesized relationships in their conceptual models. 
    \item \considerPossibilities: Analysts should have support
    in expressing ambiguous relationships and then considering multiple possible conceptual structures.
\end{itemize}

We address these goals through new language
constructs and a two-phase interactive disambiguation process in \rTisane. We also update DSL constructs to more easily specify
study design details (e.g., types of measures, syntactic sugar for specifying experimental conditions). 

%% file: rTisane.tex
\section{System Design and Implementation} \label{sec:rTisane}
\rTisaneOverview

\rTisane consists of (i) a DSL for analysts to express their conceptual models
and (ii) interactive disambiguation steps to compile this high-level
specification into a script for fitting a statistical model. A central tension in
\rTisane is how to design a usable DSL that allows statistical non-experts to
express their assumptions in a way that is still amenable to rigorous, formal
reasoning to derive statistical models. 
\autoref{fig:rTisaneOverview} gives an overview of the \rTisane system.


\def\Participant{\texttt{Participant}\xspace}
\def\Unit{\texttt{Unit}\xspace}
\def\Condition{\texttt{Condition}\xspace}
\def\Conditions{\texttt{Condition}s\xspace}

\rTisaneProgram

\subsection{\rTisanes Domain-Specific Language}
\rTisane provides language constructs for declaring variables, specifying a conceptual model, and querying for a
statistical model. 

\subsubsection{Declaring variables}
Analysts can express two types of variables: Units and Measures. Units represent
observational or experimental units from which analysts collect data (see line 5 in~\autoref{lst:rTisaneProgram}). 
A common unit is a participant in a study, so \rTisane provides syntactic sugar for
constructing a \Participant unit directly. \Participant is implemented as a wrapper for
declaring a \Unit.

Measures are attributes of Units collected in a \dataSet, so they are declared
through a Unit. Measures can be one of four
types: continuous, unordered categories (i.e., nominal), ordered categories
(i.e., ordinal), and counts (see lines 6-18 in~\autoref{lst:rTisaneProgram}). Analysts declare
unordered and ordered categories through the \texttt{categories} function.
Analysts can specify a variable is ordered by passing a list to the
\texttt{order} parameter. Otherwise, the variable is considered unordered.
Analysts can use \texttt{continuous} and \texttt{count} functions to declare
continuous and count Measures. \rTisane provides syntactic sugar for declaring
\Conditions, or discrete empirical interventions, as either unordered or ordered categories. 

\def\causes{\texttt{causes}\xspace}
\def\relates{\texttt{relates}\xspace}
\def\when{\texttt{when}\xspace}
\def\then{\texttt{then}\xspace}
\def\assume{\texttt{assume}\xspace}
\def\hypothesize{\texttt{hypothesize}\xspace}

\subsubsection{Specifying a conceptual model}
Once analysts have constructed variables, they can specify how these variables
relate conceptually. To do so, they construct a \texttt{ConceptualModel} and add
variable relationships to it (lines 20-31 in~\autoref{lst:rTisaneProgram}). The conceptual model 
is represented as a graph with variables as nodes and relationships
as edges. 

There are two types of relationships: \causes and \relates. \causes indicates a
unidirectional influence from a cause to an effect. \causes introduces a
directed edge from the cause node to the effect node. \relates indicates that
two variables are related but exactly how remains ambiguous. Analysts may be
uncertain about the direction of influence. Therefore, \relates introduces a
bi-directional edge  between two variables. During a disambiguation step,
\rTisane will walk analysts through possible graphical structures that a
bi-directional edge could represent (\considerPossibilities). To derive a
statistical model, \rTisane requires an analyst to assume a direction of
influence.

Furthermore, towards the design goal of \optionalSpecificity, \rTisane allows analysts to
optionally specify \when and \then parameters in the \causes and \relates
functions. There are four comparisons analysts can specify in
\when and \then: \texttt{increases} (for continuous, ordered categories,
counts), \texttt{decreases} (for continuous, ordered categories, counts),
\texttt{equals} (for any measure type), and \texttt{notEquals} (for any measure
type). Supporting optional specificity is designed to make the \rTisane
program an accurate document of analysts' implicit assumptions.


To add relationships to the conceptual model, analysts must \assume or
\hypothesize a relationship (\assumeHypothesize). This distinction supports analysts in distinguishing
between assumed, or strongly held, and hypothesized, or more uncertain,
relationships. 
The distinction between \assume and \hypothesize, combined with the constructs for optional
specificity, addresses analysts' inclination towards informal descriptions of
variable relationships (e.g., ``influences'') observed in the exploratory study (\autoref{subsubsec:knownVsuspected}).

Analysts can also specify interactions between two or more variables by adding
\texttt{interacts} annotations to the \texttt{ConceptualModel} (lines 30-31
in~\autoref{lst:rTisaneProgram}). Interactions provide additional information
about existing relationships in the conceptual model
(\interactionAsPartialSpec). Interactions are not distinct relationships and so
are added to the graph without \assume or \hypothesize statements. 

\def\query{\texttt{query}\xspace}
\subsubsection{Querying for a statistical model}
Analysts \query \rTisane for a statistical model based on the input conceptual
model (lines 33-34 in~\autoref{lst:rTisaneProgram}). The query asks for a
statistical model to accurately estimate the average causal effect (ACE) of the
independent variable on the dependent variable. The querying process initiates
the interactive disambiguation process, after which an \texttt{R} script
specifying and fitting a generalized linear model is output. 

\subsection{Two-step Interactive Disambiguation}
There are two phases to compiling a conceptual model to a
statistical model: (i) conceptual model refinement and (ii) statistical
model derivation. 

\subsubsection{Conceptual Model Refinement} \label{subsec:conceptualModelDisambig} 
\conceptualModelDisambiguation
The goal of the conceptual model refinement step is to make analysts' expressed
conceptual models precise enough to derive a statistical model. Conceptual model
refinement involves breaking cycles in the conceptual model by (i) picking a
direction for any \relates relationships and/or (ii) removing edges. Cycles must be broken because they imply multiple different data generating
processes that could lead to different statistical models. In this way,
conceptual model refinement can help analysts reflect on and clarify their
implicit assumptions. 

To disambiguate conceptual models, \rTisane uses a GUI.~\autoref{fig:figureConceptualModelsDisambiguation} shows the conceptual model disambiguation interface for the input program in~\autoref{lst:rTisaneProgram}. The GUI shows a graph
representing analysts' conceptual models. If there are any \relates
relationships, \rTisane suggests ways analysts could assume a direction of
influence. Additionally, \rTisane suggests ways to break any cycles in the
conceptual model. 
\rTisane finds cycles by iteratively searching for cycles of increasingly larger sizes up to the total number of nodes in the underlying graph representation. 
This algorithm takes exponential time and does not scale up to arbitrarily large graphs. 
\rTisane suggests edges in the cycle to remove in no particular order.
As analysts make changes, the graph visualization updates. The GUI
also explains why these are necessary steps to derive a statistical model. 

Once analysts have refined their conceptual models, \rTisane updates the
internal graph representation and derives a space of possible statistical
models. To narrow this space of possible statistical models down to one output
statistical model, \rTisane asks additional follow-up disambiguating questions. 

\subsubsection{Statistical model derivation and disambiguation}
\statisticalModelDisambiguation 
To formulate possible statistical models,
\rTisane considers potential covariates to control for confounding,
interactions, and family and link functions. \rTisane is able to do this because
it represents the conceptual model as a graph internally. \rTisane treats these
graphs as causal diagrams, allowing for formal reasoning about statistical model formulation. 

\paragraph{Confounder selection.} To determine confounders, \rTisane uses recent
recommendations from Cinelli, Forney, and
Pearl~\cite{cinelli2020controls}\footnote{\tisane relied on Vanderweele's
recommendations for confounder
selection~\cite{vanderweele2019modifiedDisjunctiveCriterion}, but in \rTisane we
opt for more recent recommendations.}. Cinelli et al.'s recommendations are
based on a meta-analysis of studies examining the impact of confounder selection
from graphical structures on statistical modeling accuracy. By following
Cinelli et al.'s recommendations, \rTisane includes confounders that help assess
the average causal effect of the query's independent variable on the dependent
variable as accurately as possible. 

\paragraph{Interaction term inclusion} Because interactions are treated as
partial specifications (\interactionAsPartialSpec), \rTisane searches for
interaction annotations in conceptual models. \rTisane suggests any
involving the query's dependent variable. Otherwise, \rTisane does not consider
any interactions.

\paragraph{Family and link function selection} \rTisane determines family and link functions based on the query's dependent
variable data type. For queries involving continuous dependent variables,
\rTisane considers Gaussian, Inverse Gaussian, and Gamma families. For counts,
\rTisane considers Poisson and Negative Binomial families. For ordered
categories, \rTisane considers Binomial, Multinomial, Gaussian, Inverse
Gaussian, and Gamma family functions. For unordered categories, \rTisane
considers Binomial and Multinomial family functions. \rTisane outputs
statistical models fit using the \lme package in \texttt{R}, so \rTisane
considers any family and link function combinations supported in \lme.

To inform analysts of statistical modeling choices, \rTisane shows a GUI
explaining confounder, interaction, and family and link function choices. In
addition, for more skilled analysts, \rTisane offers the opportunity to
remove any confounders or interactions based on their domain knowledge or prior
experience. Additionally, analysts must also pick a family and link function
pair if multiple possibilities could apply.
\autoref{fig:figureStatisticalModelsDisambiguation} shows the GUI for
statistical model disambiguation. 

%% file: lab-study.tex
\section{Evaluation: Controlled lab study} \label{sec:summativeEval}

Two research questions motivated our evaluation of \rTisane:

\begin{itemize}
    \item \evalConceptualModels What is the impact of \rTisane on conceptual
    modeling? Specifically, do analysts find it easy to externalize their conceptual models with \rTisane? Does \rTisane help analysts determine what implicit assumptions to specify? 
    \item \evalStatisticalModels How does \rTisane impact the statistical models
    analysts implement? Specifically, how aligned are statistical models to the conceptual models they are supposed to encode?
    How well do the statistical models analysts author on their own vs. with \rTisane fit the data? 
\end{itemize}

We focused on comparing \rTisane to a lack of tool support since the core
motivations of this work are (i) to understand how to support conceptual model
externalization and (ii) to assess the impact of externalizing conceptual
models. To our knowledge, no equivalent evaluation of \tisane has been performed. 

\subsection{Study Design}

We conducted a within-subjects (\rTisane vs. no tool support) think-aloud
lab study that consisted of four phases. 
All participants completed the phases in the
following order.

\begin{itemize}
    \item \textbf{Phase 1: Warm up} We presented participants with the
    following open-ended research question: ``What aspects of an adult's
    background and demographics are associated with income?'' We asked
    participants to specify a conceptual model including variables they thought
    influenced income. This warm-up exercise helped to externalize and keep
    track of participants' pre-conceived notions and assumptions prior to seeing
    a more restricted data schema.
    \item \textbf{Phase 2: Express conceptual models} We presented participants
    with a data schema describing a dataset from the U.S. Census Bureau. We then
    asked participants to specify a conceptual model using only the available
    variables. At the end, we asked participants about their
    experiences specifying their conceptual models in a brief survey and semi-structured interview.
    \item \textbf{Phase 3: Implement statistical models} We asked participants
    to implement ``a statistical model that assesses the influence of variables
    [they] believe to be important (in the context of additional potentially
    influential factors) on income,'' relying on only their conceptual model. 
    We then asked participants about their experiences implementing statistical
    models through a brief survey and semi-structured interview. 
    \item \textbf{Phase 4: Exit interview} The study concluded with a survey
    and semi-structured interview where we asked participants about their
    experience in the study, using \rTisane, and connecting
    conceptual models to statistical models.
\end{itemize} 

In order to assess the effect of tool support on conceptual models and the
quality of statistical models, we counterbalanced the order of tool support, or
if participants completed each task with or without rTisane first. The order of
tool use was the same for Phases 2 and 3. Specifically, within Phases 2 and
3, half the participants completed the task on their own and then with \rTisane. The
other half started with \rTisane and then did the task on their own. Prior to
using \rTisane in Phases 2 and 3, participants followed a tutorial introducing
the relevant language constructs for each task. \autoref{fig:evaluationFlow}
summarizes the evaluation's study design.

\evaluationFlow

In effect, the study compares \rTisane to a scaffolded workflow. We chose this
baseline for three reasons. First, we assume that conceptual modeling is a
helpful strategy when specifying statistical models. Second, \rTisane is
designed to both scaffold a modeling process and provide tool support for
externalizing conceptual models. Third, we wanted to isolate the effect of tool
support for externalizing conceptual models rather than measure the effect of
scaffolding plus tool support. Therefore, we anticipate that any impact of
\rTisane we observe will be more pronounced when comparing \rTisane to an
open-ended, unscaffolded analysis approach. 


\noindent \paragraph{Participants} We recruited 13 data analysts on Upwork. We
screened for participants who reported having experience with authoring
generalized linear models and using R at a three or higher on a five-point
scale. Participants self-rated their data analysis experience at a median of eight out of ten (min: 5, max: 10).~\autoref{tab:summativeEvaluationParticipants} summarizes the
participants' backgrounds. All studies were conducted over Zoom. Participants used \rTisane
on a remote controlled computer, so they did not have to install it on their
own. Each study lasted between two and three hours. Each participant was compensated
\$25 per hour. We recorded participants' screens, video, and audio throughout
the study. We then transcribed the audio.

\tableSummativeEvalParticipants

\subsection{Analysis Approach}
Our analysis procedure consisted of two parts: (i) a thematic analysis of lab
notes, transcripts, and open-ended survey questions and (ii) an artifact
analysis of conceptual models and statistical models authored with and without
\rTisane. For the conceptual models, we compared their form and content between
tool support conditions. For the statistical models, we compared the overall
statistical approach, specific statistical model formulations, rationale for
analysis decisions, and two goodness of fit measures: AIC and BIC. The first two
authors initially iterated on the thematic analysis and artifact analysis
separately. Then, they jointly revisited both and interpreted emergent
observations across the two analyses. 

One of the 13 participants dropped out part way through the study due to
discomfort with programming in front of the researchers. We analyzed the data we
were able to collect from them. 
\subsection{RQ1 Findings: \rTisanes Impact on Conceptual Model Specification}
\conceptualModelsScaffold

\textbf{Key takeaway: \rTisane scaffolded and productively constrained how analysts expressed
their conceptual models. As a result, analysts reflected on implicit domain
assumptions more deeply, considered new relationships, and felt they
accurately externalized their implicit assumptions.}

The conceptual models analysts expressed on their own were diverse in form,
content, and complexity. The majority [P2, P4, P5, P8, P11, P13] invoked a
graph-like structure. [P2, P4, P8 used \rTisane second; P5, P11, P13 used
\rTisane first].~\autoref{fig:figureConceptualModelsScaffold} illustrates four
example conceptual models from participants\footnote{An example conceptual model
given in the task instructions may have biased analysts towards a graphical
structure.}. Participants also described their conceptual models verbally [P10],
in natural language text [P6, P9], and as a timeline [P12]. As shown in
\autoref{fig:figureConceptualModelsScaffold}, P7, who used \rTisane first, even
jumped to expressing their conceptual model in a statistical model. P12's
conceptual model was particularly creative. His timeline featured variables
ordered starting on the left by how much an individual could intervene upon them
(see \autoref{fig:figureConceptualModelsScaffold}). P12's conceptual model
reiterates our finding from the exploratory lab study that analysts want to
capture nuances in a conceptual model. Furthermore, ten participants involved
all five independent variables from the \dataSet in their conceptual models [P2,
P3, P4, P5, P7, P8, P9, P11, P12, P13]. Two participants [P7, P13] also included
interactions between variables in their conceptual models. For instance, P13
specified a complex conceptual model where age, race, and sex interacted to
cause an interaction between education and employment, which then causes income (see \autoref{fig:figureConceptualModelsScaffold}).

\theme{Without \rTisane, analysts find it difficult to fully express their assumptions.}
In a survey and interview about their conceptual modeling experiences,
participants shared that they found it difficult to author conceptual models
without tool support due to doubts about how to communicate nuances in
relationships [P3, 13] and concerns about mis-specifying relationships beyond
their domain knowledge [P5, P10]. P13 explained how they wanted to
\shortquote{[i]dentify how I may weigh certain variables based on my general
awareness and knowledge and overall weights of each variable of how one may
affect income more or less in various circumstances.} Similarly, P8 described
the process of specifying their conceptual model as a general \shortquote{struggle} because \shortquote{When doing
it myself, there are so many possibilities [of expression].} While \rTisane is
not designed to prevent mis-specifications due to limited domain knowledge, we
found that \rTisanes formalism removed the need for analysts to come up with
how to express their domain knowledge. Instead, analysts could focus on expressing what they
knew. 

\theme{\rTisane encourages analysts to think about their domains more deeply.}
\rTisanes DSL deepened participants’ thinking [P3, P4, P7, P8, P10, P12, P13],
giving them, as P12 described, a structure to explore the \shortquote{boundaries of their
domain knowledge.} P3 explained how even after specifying conceptual models on
her own, \rTisanes four composable relationships (\assume, \hypothesize \texttt{x}
\causes, \relates) facilitated a deeper consideration of each relationship and
what she knew about each: \longquote{Having to think about specifics like `Do we know the
direction of the relationship' or `What happens when a category
increases/decreases' actually helped me put my thoughts out more clearly. I was
able to think about more possible scenarios that could conflict with my current
assumption, which I was probably not doing [before without rTisane]...In conclusion, I want to
say that looking at four possible ways to write a relationship made me think
more about each one of them.} Similarly, P4 explained how the DSL’s support for
optional specificity \shortquote{encouraged [them] to think about the directionality of my
hypothesized relationships and for categorical variables to think about the
effect of each individual category.}

Three participants expressed identical conceptual models with and without
\rTisane [P9, P11, P12]. Interestingly, for six participants, the
conceptual models they authored with \rTisane were subgraphs of conceptual models
authored without \rTisane[P2, P3, P4, P5, P7, P8]. 
For P2, P3, P4, and P8, who used \rTisane second, \rTisane appeared to help
focus them on the specifics of variables and relationships of interest. P4 explained, \shortquote{coding made it [the conceptual model] more specific}.
On the other hand, P5 and P7, both of whom used \rTisane first, expanded upon conceptual models specified with \rTisane when asked to
subsequently express conceptual models on their own. For example, P7 authored a
statistical model involving an interaction between variables in their \rTisane
conceptual model when asked to specify a conceptual model on their own. It seems
that just conceptual modeling with \rTisane helped P7 translate a conceptual
model to a statistical model on his own. Taking these observations together, we
see that \rTisanes DSL can support both convergent and divergent creative
thinking about analysts' domain knowledge. 

\theme{\rTisane provides structure to express conceptual models easily and accurately.}
Participants appreciated how \rTisane structured their conceptual modeling
process [P2, P4, P9, P10, P11, P12, P13]. 
Four participants said
that \rTisane generally made it easier for them to specify their conceptual
models [P4, P8, P10, P12]. P4 and P10 even believed that rTisane’s \shortquote{formal
structure made [conceptual modeling] more rigorous} [P4] and \shortquote{more
disciplined} [P10]. 
P10 continued, 
\longquote{My thinking
was that before I didn’t have much idea about how can I link my variable with
the output [variable], and how this can interact. And so it may need some trial
and error... using this API, there are predefined functions, they are translated
in R language, cause or relates, it made my task easier. This translation was
not on me anymore.} 

Participants relied on the conceptual disambiguation step to verify that what
they expressed in code accurately represented their implicit assumptions [P2,
P8, P12]. P2, who had drawn a conceptual model as a graph on his own prior to
using \rTisane, said, \shortquote{The interactive process was a good
way to check that the graph came out the same way I was picturing it. It was
helpful because it is easier to look at than code.}
\rTisanes specify-then-refine approach to expressing conceptual models helped
analysts feel confident that they expressed their conceptual assumptions
accurately.

\subsection{RQ2 Findings: \rTisanes Impact on Statistical Model Quality}
\tableModelScores
\textbf{Key takeaway: 
\rTisane focused participants on their analysis goals over low-level details
that bogged them down without tool support. 
As a result, \rTisane helped analysts maintain their analysis intent during statistical modeling.
\rTisane also improved the
statistical model authoring process and how well the output statistical models fit the data.
}

On their own, three participants were not able to author a statistical model due
to unfamiliarity with statistical methods [P3], lack of time [P5], and reliance
on visual analyses (ie.g., heatmaps, scatterplots) [P12]. Of the remaining nine
participants who completed the study, six participants successfully authored
linear regression models [P2, P4, P7, P8, P9, P10]. A seventh participant, P6,
started to author a logistic regression model with Race and Income but stopped
before binarizing Income. Two participants, both of whom had just
finished authoring statistical models with rTisane, implemented GLMs [P11, P13].
P11 based their own statistical model (in the no tool support condition) on the \rTisane output model script. 
After observing the model's \shortquote{AIC is large, the residual is
large,} P11 determined \shortquote{I don't think this [\rTisane output model] is the right
fit.} So, they log-transformed the income variable and fit a new statistical model. 
P11’s experience mirrors how we anticipate analysts will iterate on
\rTisanes output statistical models in the future.

\theme{Without \rTisane, analysts change their analysis intent during statistical modeling.} \label{themeChangeAnalysisIntent}
Without \rTisane, participants [P2, P5, P6, P8, P10], adopted a more exploratory
or data-focused approach, changing their analysis goals while authoring
statistical models. This theme is best illustrated by P2, who started with a
hypothesis that Occupation, or Employment, influenced Income. His conceptual
model in \rTisane had the variables Education, Age, Race, and Sex causing
Occupation (Employment), which in turn, causes Income (see \autoref{fig:figureConceptualModelsScaffold}). 

He started authoring statistical models with the intent to assess this
hypothesis. On his own, he first authored an ANOVA with Employment as the IV and
Income as the DV. Once he saw that Employment had a statistically significant
influence on Income, he changed his analysis goal to assessing if the variables
causing Employment would \shortquote{be able to predict which occupation [employment]...And then...the
income from the occupation [employment] just because that’s how I like structured it [in the
conceptual model] initially.} However, P2 got stuck on how to author a model
with Employment as the outcome variable because it was categorical, saying,
\shortquote{But the way I structured it in like the diagram. I'm not sure exactly how to
do that, because Occupation's [Employment's] like categorical. Um, so I'm not sure like
exactly...how to model that.} This roadblock led P2 to consider an alternative
\shortquote{regression model with Income as like the output and then...all [the IVs] as
terms and then just include the interactions between Occupation [Employment] and the terms
that were pointing into it, and that would just be one model.} In other words,
P2 tried to author a single statistical model to assess if there was evidence
for his conceptual model. However, he was unaware of three key things. First,
given his conceptual model, he did not need to account for the other variables
to estimate the influence of Employment on Income and assess his hypothesis.
Second, adding interaction terms would not capture the dependencies in the
conceptual model. Third, P2 likely needs a structural equation model to assess all the relationships in his conceptual model.

While it is well documented that statistical analysis is an iterative
process~\cite{grolemund2014cognitive, jun2022hypoForm} and we saw evidence of
this among participants [P5, P6, P10, P11, P12], what P2's experience
exemplifies is how creative participants can be in convincing themselves that the
statistical model they authored not only assessed a particular hypothesis but
could also arbitrate if their entire conceptual models were supported by data.
Furthermore, this suggests an opportunity for \rTisane to support a more
iterative analysis process and help analysts author multiple models to assess an
entire conceptual model, not just the influence of a single independent variable
on a dependent variable, and idea we expand upon in \autoref{sec:limitations}.




\theme{\rTisane focuses analysts on their motivations for analysis.}
In contrast, participants reported that \rTisane guided them to think about
their domains more [P2, P12], lightened their burden in authoring statistical
models [P10], and even promoted research transparency [P5] and reproducibility
[P4]. Furthermore, rTisane reinforced prior knowledge about statistical methods
[P6, P11] and helped participants learn more about GLMs [P4, P6, P7, P13]. P6,
who had tried to author a logistic regression model on her own, explained how
she could apply what she learned from using \rTisane to future analyses: \shortquote{I
like that a multivariate linear regression was used because this will inform
any future data analysis...}

\theme{Without \rTisane, analysts find statistical model formulation challenging.}
Participants reported formulating and evaluating statistical models [P2, P3, P5,
P8, P12], programming [P6, P13], and preparing data [P7] as the major challenges
to authoring statistical models without rTisane. For example, P3 explained how
\longquote{There are a number of statistical tests and it gets confusing if I don't
practice it frequently. This is what happened today, I haven't worked on a
hypothesis testing problem recently and while I knew what libraries to go to, I
was not sure which test to implement.} Similarly, discussing the details of
which covariates to include in a statistical model given a conceptual model, P8
explained how he was uncertain about which \shortquote{upstream relationships,} or
indirect causes, to include in a statistical model. Without \rTisane, he
described statistical model authoring as \shortquote{It immediately feels harder doing it
directly [without rTisane] like this} [P8].

\theme{Statistical models authored with \rTisane fit the data just as well or better than statistical models without \rTisane.}
Of the eight participants who successfully authored linear regression or
GLMs on their own, three implemented identical models with
or without \rTisane [P7, P9, P13]. Notably, all three had authored the
statistical model with \rTisane first, suggesting that \rTisane biased their own
modeling processes. \autoref{tab:statisticalModelComparison} shows statistical models authored with and without \rTisane and their AIC and BIC goodness of fit measures. For another three participants [P4, P8, P10], their
statistical models with \rTisane had lower AIC and BIC scores than the
statistical models without \rTisane. In other words, for P4, P8, and P10,
\rTisane helped them author statistical models that fit the data better. Thus,
for six out of eight participants, \rTisanes statistical models fit the data
better or equally well. For P11, the statistical model they authored without
\rTisane dropped some observations, so the models are not directly comparable.
For P2, the \rTisane statistical model fit worse than his own statistical model
in part due to an observed change in his motivation for analysis, discussed
above (\autoref{themeChangeAnalysisIntent}). 



\subsection{Opportunities to Improve \rTisane}
While participants found \rTisane helpful, they suggested two areas of
tool improvement: (i) family and link function selection and (ii) statistical model
interpretation. Additionally, participants expressed wanting to use \rTisane for
scientific communication, not just statistical authoring. These ideas require
future research and have the potential to help analysts engage with and
understand their analyses more deeply.

Several participants had difficulty
picking family and link functions [P2, P4, P5, P9, P10, P11]. P4 explained,
\shortquote{I didn't understand the benefit or tradeoffs between different
specifications. It wasn't obvious to me how to create a linear OLS regression,
or why I would want to use a specification besides linear OLS.} This problem arises from the 
stark contrast between \rTisanes
relatively high-level conceptual modeling abstractions, and \rTisanes statistical disambiguation step that requires
analysts to select specific family and link functions, a relatively
low-level statistical modeling detail. 
Therefore, an important next step is to incorporate approaches to suggest a specific pair
of family and link functions and interfaces that explain the
\shortquote{tradeoffs} in choices. 

Because \rTisane uses lme4 under the hood, the result of executing the output
statistical modeling script is the output from lme4. However, analysts expected
the outputs to at least relate back to their conceptual models, given that
\rTisanes DSL is focused on conceptual modeling. For example, P8 found the
output from lme4 overwhelming, saying, \shortquote{Looking at the
\texttt{summary()} in R was too much to look at.} He suggested a simple way to
tie the results back to his input conceptual model: \shortquote{Would be nice if
you could have the same visual representation with p-values/coefficients!}
Future work should explore ways to make statistical modeling
output more understandable for statistical non-experts. 

Lastly, when asked how they might imagine using \rTisane,
participants described how experienced and novice analysts
alike would benefit from using \rTisane [P2, P4, P9, P10, P12]. 
Participants also suggested that conceptual models written using \rTisane could help collaborations with less technical
stakeholders [P8, P9]. For instance, P8 detailed how a conceptual model written using rTisane
could be a communication tool, saying how the \shortquote{visual representation would play
a role in a dialogue with the PI.} P8 went on to imagine how he would like to
use \rTisanes conceptual model to generate process diagrams in scientific
papers. We expand upon this possibility in \autoref{sec:limitations}.

%% file: discussion.tex
\section{Discussion}
This work has demonstrated how externalizing conceptual models (i) increases
engagement with implicit assumptions and (ii) leads to higher quality
statistical models that correspond to conceptual models. \rTisane structures a
conceptual model specification process that deepens reflection by providing a
usable DSL and interactive disambiguation process for refining a conceptual
model after initial specification. \rTisane also guarantees fidelity between
conceptual and statistical models by translating expressed conceptual models
into causal diagrams to inform statistical model formulation. In the evaluative
study, we find that the statistical models analysts author with \rTisane answer
their motivating research questions. 
Additionally, statistical models authored with \rTisane, in many cases, 
fit real-world data better than statistical models authored without \rTisane. 
In other cases, \rTisane's output statistical models serve as the basis for further model tuning. 
These findings demonstrate the benefits of formalism, designing for both
usability and rigor in DSLs, and the potential for shared representations to
become boundary objects. 

While interfaces leveraging natural language, especially in the era of large
language models and their applications (e.g., ChatGPT~\cite{brown2020language}),
are enticing, we find that analysts in our evaluation preferred the structure
provided by \rTisanes DSL over open-ended specification without \rTisane.
\rTisane focused analysts on \textit{what} to express about their implicit
assumptions while also providing them with easy-to-learn syntax for doing so.
Analysts used \rTisanes DSL as starting points to dig deeper into their domain
knowledge, such as distinguishing assumptions based on prior literature from
their own hypotheses, and to consider new relationships. As a result, analysts'
reported that the conceptual models expressed with \rTisane accurately
represented their internalized knowledge. In other words, \rTisanes formalism
promoted what Donald Schön calls
\textit{reflection-in-action}~\cite{schon1987reflective} during data analysis.
Therefore, HCI researchers should consider how formalisms can facilitate a
sensemaking process~\cite{russell1993cost} that helps users attain their
ultimate goal. 

DSLs need to be both usable for people to write programs in them and rigorously designed for
automation to accomplish specific tasks. 
To ensure usability of \rTisanes DSL, we iteratively design language constructs
and disambiguating interactions. We use an existing DSL to probe into what and
how analysts want to express their implicit knowledge, design \rTisane, and
evaluate it in a controlled lab study. We design for rigor in the compilation
process from an input conceptual model specification to a statistical model
representing the conceptual model and analysis intent. In the evaluation, we
find that analysts could easily translate their domain knowledge into conceptual
models using the \rTisane DSL (usability), which then generated statistical
models that addressed the motivating research question and fit the data better
than hand-coded statistical models (rigor). In this way, \rTisane exemplifies
the synergy of usability and rigor by leveraging the conceptual model as a
shared representation~\cite{heer2019agency}.

In the evaluation, participants discussed the potential for using conceptual
models to communicate assumptions and analyses with collaborators and the
broader scientific community. Specifically, participants mentioned the value of
conceptual models as a record of the analyst's thoughts for future analysts, as
a way of summarizing the analysis for less technical collaborators, and as a way
of generating process diagrams for scientific papers.  In all of these
applications, conceptual models serve as an \textit{intermediate representation}
that can be ``compiled'' to a number of ``backends.'' Furthermore, these
applications suggest that conceptual models are likely useful as boundary
objects~\cite{star1989boundaryObjects} for collaboration and communication, a
future research direction worth pursuing. In this light, \rTisane is just one
example application for how to leverage conceptual models to increase scientific
and statistical literacy, transparency, and reproducibility. 


Finally, to our knowledge, the evaluation is the first to explicitly capture
analysis intent and observe analysts exhibiting risky conceptual drift in their
standalone statistical models, a phenomenon not seen when using formal model
specification and generation. This suggests that focusing on low-level
programming details may impede analysts from authoring statistical models that
represent implicit assumptions and address analysis intent. We anticipate this
contrast would be even starker if we had asked analysts to author statistical
models on their own and did not guide them through the scaffolded workflow
without \rTisane. 


%% file: limitations.tex
\section{Limitations and Future Work} \label{sec:limitations}

There are three important limitations to this work that present opportunities for future work. 

First and foremost, our goal has been to investigate (i) how to support analysts in
externalizing their conceptual models so that they accurately represent
analysts' implicit assumptions and (ii) the impact of this process on
statistical modeling. 
To answer these questions, we focus on the design of a DSL because both analysts
and analysis tool developers can leverage DSLs. While we believe that other
interactive tools for conceptual modeling would benefit from incorporating our
insights from the design and evaluation of \rTisane, precisely how to do so
remains an avenue for future work. For instance, could drawing-based tools like
Dagitty's web interface~\cite{textor2011dagitty} provide analysts with drop-down
menu options for labeling conceptual relationships as known or suspected? How
would these designs affect the conceptual models analysts express? Could these
designs make existing interactive tools for externalizing conceptual models more
usable for statistical non-experts? Future work should address these questions. 

Second, there are two important limitations to the lab evaluation: the number
and backgrounds of participants. Our sample size of 13 is limited. Nevertheless,
we reached convergence and saturation of themes while analyzing transcripts and
researcher notes. Moreover, we recruited participants through the online
freelance platform Upwork. As a result, our participants came from a variety of
disciplines and were data analysis practitioners and educators
(\autoref{tab:summativeEvaluationParticipants}). 
We filtered for participants
who self-reported familiarity with generalized linear modeling and R. However,
some struggled with R syntax, suggesting that their self-reported skills were
inflated. Therefore, it seems that \rTisane is able to help even those with less
expertise than we expected. 
A similar limitation exists for our exploratory study involving CS PhD students
in a research methods course.
Future work should focus on assessing the impact of
\rTisane on novice analysts from specific disciplines in order to reveal additional language constructs or
interactions to help a wider range of users. 







Finally, we believe future tool support for statistical model iteration is
crucial. Using \rTisane, participants could iterate on their conceptual models
by adding or removing variables and relationships, but they could not engage in
a larger iteration loop with their output statistical model. 
For instance, P11 described the \rTisane output statistical model as
\shortquote{an initial or baseline model but follow-up evaluation of the model
is needed.} They wanted to \shortquote{go back and tweak things a bit} about
their statistical model. Tools like \rTisane should ensure that analysts
maintain their analysis intents throughout iteration---or at least document
conceptual shifts---while discouraging or even preventing analysts from
questionable ``data dredging'' or HARKing~\cite{kerr1998harking} practices. A
first step may be to support recommended workflows for statistical model
development and refinement, such as Gelman et al.'s Bayesian
Workflow~\cite{gelman2020modelExpansion}.

%% file: conclusion.tex
\section{Conclusion}
\rTisane provides a DSL with language constructs for expressing conceptual
models and integrates a two-phase interactive
disambiguation process for compiling conceptual knowledge into statistical
analysis code. In a controlled lab study of \rTisane, we
find that the DSL is expressive enough to capture analysts' conceptual models
accurately, eases the burden of making their implicit assumptions explicit, and
pushes analysts to think about their domains more deeply. Using \rTisane,
analysts are able to author statistical models that fit the data just as well
as, if not \textit{better} than, statistical models authored on their own. 
Analysts also report that through the process,
they learn about GLMs. Together, these results
demonstrate how 
externalizing conceptual models during data analysis 
enables statistical non-experts to author high-quality statistical models that they would not be
able to author otherwise.

%% file: summative-eval-paper.bbl

\begin{thebibliography}{00}


\ifx \showCODEN    \undefined \def \showCODEN     #1{\unskip}     \fi
\ifx \showDOI      \undefined \def \showDOI       #1{{\tt DOI:}\penalty0{#1}\ }
  \fi
\ifx \showISBNx    \undefined \def \showISBNx     #1{\unskip}     \fi
\ifx \showISBNxiii \undefined \def \showISBNxiii  #1{\unskip}     \fi
\ifx \showISSN     \undefined \def \showISSN      #1{\unskip}     \fi
\ifx \showLCCN     \undefined \def \showLCCN      #1{\unskip}     \fi
\ifx \shownote     \undefined \def \shownote      #1{#1}          \fi
\ifx \showarticletitle \undefined \def \showarticletitle #1{#1}   \fi
\ifx \showURL      \undefined \def \showURL       #1{#1}          \fi

\bibitem{alspaugh2018futzing}
{Sara Alspaugh}, {Nava Zokaei}, {Andrea Liu}, {Cindy Jin}, {and} {Marti~A
  Hearst}. 2018.
\newblock \showarticletitle{Futzing and moseying: Interviews with professional
  data analysts on exploration practices}.
\newblock {\em IEEE transactions on visualization and computer graphics\/}
  {25}, 1 (2018), 22--31.
\newblock


\bibitem{causalWizard}
{AI/ML~Services (Australia)}. 2023.
\newblock CausalWizard.
\newblock   (2023).
\newblock
\showURL{%
\url{"https://causalwizard.app/"}}


\bibitem{battle2019EVA}
{Leilani Battle} {and} {Jeffrey Heer}. 2019.
\newblock \showarticletitle{Characterizing Exploratory Visual Analysis: A
  Literature Review and Evaluation of Analytic Provenance in Tableau}.
\newblock {\em Computer Graphics Forum (Proc. EuroVis)\/} (2019).
\newblock
\showURL{%
\url{"http://idl.cs.washington.edu/papers/exploratory-visual-analysis"}}


\bibitem{baughan2022dissociation}
{Amanda Baughan}, {Mingrui~Ray Zhang}, {Raveena Rao}, {Kai Lukoff}, {Anastasia
  Schaadhardt}, {Lisa~D Butler}, {and} {Alexis Hiniker}. 2022.
\newblock \showarticletitle{“I Don’t Even Remember What I Read”: How
  Design Influences Dissociation on Social Media}. In {\em Proceedings of the
  2022 CHI Conference on Human Factors in Computing Systems}. 1--13.
\newblock


\bibitem{bezanson2017julia}
{Jeff Bezanson}, {Alan Edelman}, {Stefan Karpinski}, {and} {Viral~B Shah}.
  2017.
\newblock \showarticletitle{Julia: A fresh approach to numerical computing}.
\newblock {\em SIAM review\/} {59}, 1 (2017), 65--98.
\newblock


\bibitem{brown2020language}
{Tom Brown}, {Benjamin Mann}, {Nick Ryder}, {Melanie Subbiah}, {Jared~D
  Kaplan}, {Prafulla Dhariwal}, {Arvind Neelakantan}, {Pranav Shyam}, {Girish
  Sastry}, {Amanda Askell}, {and} {others}. 2020.
\newblock \showarticletitle{Language models are few-shot learners}.
\newblock {\em Advances in neural information processing systems\/}  {33}
  (2020), 1877--1901.
\newblock


\bibitem{cinelli2020controls}
{Carlos Cinelli}, {Andrew Forney}, {and} {Judea Pearl}. 2020.
\newblock \showarticletitle{A crash course in good and bad controls}.
\newblock {\em Sociological Methods \& Research\/} (2020), 00491241221099552.
\newblock


\bibitem{gelman2020modelExpansion}
{Andrew Gelman}, {Aki Vehtari}, {Daniel Simpson}, {Charles~C Margossian}, {Bob
  Carpenter}, {Yuling Yao}, {Lauren Kennedy}, {Jonah Gabry}, {Paul-Christian
  B{\"u}rkner}, {and} {Martin Modr{\'a}k}. 2020.
\newblock \showarticletitle{Bayesian workflow}.
\newblock {\em arXiv preprint arXiv:2011.01808\/} (2020).
\newblock


\bibitem{grolemund2014cognitive}
{Garrett Grolemund} {and} {Hadley Wickham}. 2014.
\newblock \showarticletitle{A cognitive interpretation of data analysis}.
\newblock {\em International Statistical Review\/} {82}, 2 (2014), 184--204.
\newblock


\bibitem{heer2019agency}
{Jeffrey Heer}. 2019.
\newblock \showarticletitle{Agency plus automation: Designing artificial
  intelligence into interactive systems}.
\newblock {\em Proceedings of the National Academy of Sciences\/} {116}, 6
  (2019), 1844--1850.
\newblock


\bibitem{sas}
{SAS~Institute Inc.} 2021.
\newblock SAS.
\newblock   (2021).
\newblock
\showURL{%
\url{https://www.sas.com/}}


\bibitem{jun2021hypothesisFormalization}
{Eunice Jun}, {Melissa Birchfield}, {Nicole De~Moura}, {Jeffrey Heer}, {and}
  {Rene Just}. 2021.
\newblock \showarticletitle{Hypothesis Formalization: Empirical Findings,
  Software Limitations, and Design Implications}. In {\em ACM Transactions on
  Computer-Human Interaction (TOCHI)}, Vol.~29.
\newblock
Issue 1.
\showURL{%
\url{"https://arxiv.org/abs/2104.02712"}}


\bibitem{jun2022hypoForm}
{Eunice Jun}, {Melissa Birchfield}, {Nicole De~Moura}, {Jeffrey Heer}, {and}
  {Rene Just}. 2022.
\newblock \showarticletitle{Hypothesis formalization: Empirical findings,
  software limitations, and design implications}.
\newblock {\em ACM Transactions on Computer-Human Interaction (TOCHI)\/} {29},
  1 (2022), 1--28.
\newblock


\bibitem{jun2019tea}
{Eunice Jun}, {Maureen Daum}, {Jared Roesch}, {Sarah~E Chasins}, {Emery~D
  Berger}, {Rene Just}, {and} {Katharina Reinecke}. 2019.
\newblock \showarticletitle{Tea: A High-level Language and Runtime System for
  Automating Statistical Analysis}. In {\em Proceedings of the 32nd Annual
  Symposium on User Interface Software and Technology}. ACM.
\newblock


\bibitem{jun2022tisane}
{Eunice Jun}, {Audrey Seo}, {Jeffrey Heer}, {and} {Ren{\'e} Just}. 2022.
\newblock \showarticletitle{Tisane: Authoring Statistical Models via Formal
  Reasoning from Conceptual and Data Relationships}.
\newblock {\em ACM CHI\/} (2022).
\newblock


\bibitem{kandel2012enterprise}
{Sean Kandel}, {Andreas Paepcke}, {Joseph~M Hellerstein}, {and} {Jeffrey Heer}.
  2012.
\newblock \showarticletitle{Enterprise data analysis and visualization: An
  interview study}.
\newblock {\em IEEE Transactions on Visualization and Computer Graphics\/}
  {18}, 12 (2012), 2917--2926.
\newblock


\bibitem{kerr1998harking}
{Norbert~L Kerr}. 1998.
\newblock \showarticletitle{HARKing: Hypothesizing after the results are
  known}.
\newblock {\em Personality and social psychology review\/} {2}, 3 (1998),
  196--217.
\newblock


\bibitem{liu2019understanding}
{Jiali Liu}, {Nadia Boukhelifa}, {and} {James~R Eagan}. 2019b.
\newblock \showarticletitle{Understanding the Role of Alternatives in Data
  Analysis Practices}.
\newblock {\em IEEE transactions on visualization and computer graphics\/}
  {26}, 1 (2019), 66--76.
\newblock


\bibitem{liu2019paths}
{Yang Liu}, {Tim Althoff}, {and} {Jeffrey Heer}. 2019a.
\newblock \showarticletitle{Paths Explored, Paths Omitted, Paths Obscured:
  Decision Points \& Selective Reporting in End-to-End Data Analysis}.
\newblock {\em arXiv preprint arXiv:1910.13602\/} (2019).
\newblock


\bibitem{mcelreath2020statistical}
{Richard McElreath}. 2020.
\newblock {\em Statistical rethinking: A Bayesian course with examples in R and
  Stan}.
\newblock CRC press.
\newblock


\bibitem{russell1993cost}
{Daniel~M Russell}, {Mark~J Stefik}, {Peter Pirolli}, {and} {Stuart~K Card}.
  1993.
\newblock \showarticletitle{The cost structure of sensemaking}. In {\em
  Proceedings of the INTERACT'93 and CHI'93 conference on Human factors in
  computing systems}. ACM, 269--276.
\newblock


\bibitem{jmp}
{SAS}. 2020.
\newblock JMP.
\newblock   (2020).
\newblock
\showURL{%
\url{"https://www.jmp.com/en_us/home.html"}}


\bibitem{schon1987reflective}
{Donald~A Sch{\"o}n}. 1987.
\newblock {\em Educating the reflective practitioner: Toward a new design for
  teaching and learning in the professions.}
\newblock Jossey-Bass.
\newblock


\bibitem{sharma2020dowhy}
{Amit Sharma} {and} {Emre Kiciman}. 2020.
\newblock \showarticletitle{DoWhy: An end-to-end library for causal inference}.
\newblock {\em arXiv preprint arXiv:2011.04216\/} (2020).
\newblock


\bibitem{spss}
{IBM SPSS}. 2021.
\newblock SPSS Software.
\newblock   (2021).
\newblock
\showURL{%
\url{https://www.ibm.com/analytics/spss-statistics-software}}


\bibitem{star1989boundaryObjects}
{Susan~Leigh Star} {and} {James~R Griesemer}. 1989.
\newblock \showarticletitle{Institutional ecology,translations' and boundary
  objects: Amateurs and professionals in Berkeley's Museum of Vertebrate
  Zoology, 1907-39}.
\newblock {\em Social studies of science\/} {19}, 3 (1989), 387--420.
\newblock


\bibitem{textor2011dagitty}
{Johannes Textor}, {Juliane Hardt}, {and} {Sven Kn{\"u}ppel}. 2011.
\newblock \showarticletitle{DAGitty: a graphical tool for analyzing causal
  diagrams}.
\newblock {\em Epidemiology\/} {22}, 5 (2011), 745.
\newblock


\bibitem{vanderweele2019modifiedDisjunctiveCriterion}
{Tyler~J VanderWeele}. 2019.
\newblock \showarticletitle{Principles of confounder selection}.
\newblock {\em European journal of epidemiology\/} {34}, 3 (2019), 211--219.
\newblock


\bibitem{winters2021heartbeat}
{R~Michael Winters}, {Bruce~N Walker}, {and} {Grace Leslie}. 2021.
\newblock \showarticletitle{Can You Hear My Heartbeat?: Hearing an Expressive
  Biosignal Elicits Empathy}. In {\em Proceedings of the 2021 CHI Conference on
  Human Factors in Computing Systems}. 1--11.
\newblock


\bibitem{wongsuphasawat2019EDAgoals}
{Kanit Wongsuphasawat}, {Yang Liu}, {and} {Jeffrey Heer}. 2019.
\newblock \showarticletitle{Goals, Process, and Challenges of Exploratory Data
  Analysis: An Interview Study}.
\newblock {\em arXiv preprint arXiv:1911.00568\/} (2019).
\newblock


\end{thebibliography}
